\documentclass[twocolumn,pra,aps,floatfix,longbibliography]{revtex4-1}
\usepackage{amsmath,esint}
\usepackage{amsfonts}
\usepackage{amssymb}
\usepackage{gensymb}
\usepackage{siunitx}
\usepackage{verbatim}
\usepackage{graphicx}
\usepackage{textcomp}
\usepackage{times}
\usepackage{natbib}
\usepackage{hyperref}
\usepackage{dsfont}
\usepackage{url}
\usepackage{appendix}

\begin{document}

\title{Quantum gyro-electric effect: Photon spin-1 quantization in continuum topological bosonic phases}

\author{Todd Van Mechelen}
\author{Zubin Jacob}
\email{zjacob@purdue.edu}
\affiliation{Birck Nanotechnology Center and Purdue Quantum Center, Department of Electrical and Computer Engineering, Purdue University, West Lafayette 47907, Indiana, USA}

\begin{abstract}

Topological phases of matter arise in distinct fermionic and bosonic flavors. The fundamental differences between them are encapsulated in their rotational symmetries - the spin. Although spin quantization is routinely encountered in fermionic topological edge states, analogous quantization for bosons has proven elusive. To this end, we develop the complete electromagnetic continuum theory characterizing 2+1D topological bosons, taking into account their intrinsic spin and orbital angular momentum degrees of freedom. We demonstrate that spatiotemporal dispersion (momentum and frequency dependence of linear response) captures the matter-mediated interactions between bosons and is a necessary ingredient for topological phases. We prove that the bulk topology of these 2+1D phases is manifested in transverse spin-1 quantization of the photon. From this insight, we predict two unique bosonic phases - one with even parity $C=\pm 2$ and one with odd $C=\pm 1$. To understand the even parity phase $C=\pm 2$, we introduce an exactly solvable model utilizing non-local optical Hall conductivity and reveal a single gapless photon at the edge. This unidirectional photon is spin-1 helically quantized, immune to backscattering, defects, and exists at the boundary of the $C=\pm 2$ bosonic phase and any interface - even vacuum. The contrasting phenomena of transverse quantization in the bulk, but longitudinal (helical) quantization on the edge is addressed as the quantum gyro-electric effect (QGEE). We also validate our bosonic Maxwell theory by direct comparison with the supersymmetric Dirac theory of fermions. To accelerate the discovery of such bosonic phases, we suggest two new probes of topological matter with broken time-reversal symmetry: momentum-resolved electron energy loss spectroscopy and cold atom near-field measurement of non-local optical Hall conductivity. 

\end{abstract}

\maketitle 

\section{Introduction}

Initial observations of topological phases of matter surfaced with the quantum Hall effect (QHE), a discovery which revealed that the transverse conductivity $\sigma_H=n\frac{e^2}{h}$ is naturally quantized  \cite{Klitzing1980,Laughlin1981,Thouless1982}. $e$ is the elementary charge of the electron and $h$ is the Planck constant. Here, $n\in \mathbb{Z}$ is the electronic Chern number and represents a global topological invariant. Being a global property of the bulk electronic band structure, it is insensitive to disorder within the material. Yet, in terms of the photon with frequency $\omega$ and momentum $k$,
\begin{equation}
\sigma_H(0,0)=n\frac{e^2}{h},
\end{equation}
only describes the local static response $\omega=k=0$ and contains no information of the high-frequency  $\omega>0$, short-wavelength $k> 0$ behavior of the electromagnetic field. The AC dynamical equivalent $\sigma_H(\omega,0)$ of the conventional DC conductivity $\sigma_H(0,0)$ is known as the optical Hall conductivity. It is measured using the Faraday rotation angle (gyrotropic response) and has shown plateau-like behavior up to THz frequencies \cite{Shimano2010}. The purpose of this paper is to unravel the global topological properties of the photon and the role of spin-1 quantization in the generalized optical Hall conductivity $\sigma_H(\omega,k)$. 

Conventionally, topological materials have focused on fermionic behavior, which display spin-\textonehalf{} polarized edge states and integer quantization of the Hall conductivity \cite{Kane2005,Bernevig1757}. However, spin-1 bosonic phases with broken time-reversal symmetry (TRS) have recently been proposed \cite{Tian2016,Chen1604,Vishwanath2013,Metlitski2013,Senthil2013,Vishwanath2012,Wen2011,van_mechelen2017} and correspond to \textbf{even} integer Hall quantization. Pioneering research in topological photonics has mimicked the fermionic behavior using carefully structured pseudo-spin-\textonehalf{} photonic crystals \cite{HafeziM.2013,Karzig2015,Alu2017,Lein2017,Zheng2015}. A few striking examples are gyrotropic photonic crystals \cite{Lu2013,Wang2009,Wang:05}, Floquet topological insulators \cite{Rechtsman2013} and bianisotropic metamaterials \cite{Khanikaev2013,Slobozhanyuk2017,He4924} which support chiral photonic edge states. Similar pseudo-spin approaches utilizing Haldane models on honeycomb lattices have led to Chern insulators \cite{Haldane2008}. These are quantum Hall phases but with zero field - realized in photonic crystals, circuit QED \cite{Houck2010} and cold atom systems \cite{Zoller2003}. Important work has also developed Chern invariants for continuous photonic media with broken TRS \cite{Silveirinha2015,Silveirinha2017,Jin2016,Song2018}. Nevertheless, the discovery of true spin-1 quantized phases has remained an open problem, as well as the connection between bosonic and photonic topologies. We solve both these problems simultaneously which can open interesting avenues for condensed matter physics and photonics.


The essential difference between fermions and bosons is revealed in their half-integer vs. integer spins. This difference is directly reflected in single-particle geometric phases \cite{BB1987,Stone2016} and arises from their rotational symmetries ($\mathcal{R}$). Under cyclic revolution, a fermion returns out of phase with itself $\mathcal{R}(2\pi)=-1$, meaning topological monopoles exhibit half-integer quantization. Conversely, bosons return in phase under the same rotation $\mathcal{R}(2\pi)=+1$, signaling integer monopoles in the band structure. Due to this critical distinction, fermions and bosons constitute different topological classes which are incommensurable with one another. Although a host of naturally occurring fermionic phases have been discovered \cite{Hasan2010}, no bosonic equivalent has been found till date. In this paper, we develop the theory of TRS broken bosonic phases for light to accelerate their discovery.

We put forth the complete microscopic continuum theory describing all 2+1D bosonic phases of the photon. We account for the inherent spin-1 symmetries of the electromagnetic field such that the bosonic properties emerge naturally. This marks a distinct departure from previous attempts at building topological field theories for the photon. We reveal that the signature of these topological bosonic phases is bulk transverse spin quantization \cite{barnett_natures_2016,BLIOKH20151,VanMechelen:16,kalhor_universal_2016} - in stark contrast to conventional photonic media where transverse spin is a continuous classical number. From very general symmetry arguments, we predict two unique photonic phases, with even $C=\pm 2$ or odd $C=\pm 1$ parity.

We show the fundamental necessity of spatiotemporal dispersion (momentum and frequency dependence of linear response) to define global topological invariants in continuum phases of matter. Spatiotemporal dispersion is a natural consequence of matter-mediated interactions between bosonic fields. We introduce an exactly solvable model, exploiting non-local optical Hall conductivity $\sigma_H(\omega,k)$, to unravel the topological physics of the even parity phase $C=\pm 2$. This phase has been predicted in interacting bosonic systems and corresponds to a single gapless photon at the edge. The unidirectional photon exists at the boundary of the non-trivial gyrotropic medium and arbitrary material interface, unlike any previously known edge states in electromagnetism. It hosts many intriguing optical properties, such as spin-1 helical quantization, anomalous displacement currents and robustness to disorder. We address the contrasting phenomena of transverse quantization in the bulk and longitudinal (helical) quantization at the edge as the quantum gyro-electric effect (QGEE). To rigorously validate our bosonic predictions, we directly compare this model to its supersymmetric Dirac theory \cite{Dunne1999,BOYANOVSKY1986483,Ziegler1998,gates2001}, highlighting the striking similarities, but important differences, between spin-1 and spin-\textonehalf{} topologies. Finally, we suggest experimental probes to search for these new bosonic phases of matter.

This article is organized as follows. In Sec.~\ref{sec:Continuum_Photonics} we analyze the linear response theory of 2+1D electromagnetism and derive the regularized continuum Hamiltonian with broken TRS. In Sec.~\ref{sec:Rotational_Symmetry} we study the rotational symmetries of this Hamiltonian and discuss the physical implications of orbital, spin and total angular momentum of the collective light-matter excitations. The following Sec.~\ref{sec:Bosonic_Phases} relates integer spin directly to the Chern number and all topological bosonic phases of the photon are found. Using an exactly solvable model, the even parity bosonic phase $C=\pm 2$ is examined extensively. Sec.~\ref{sec:Dirac_Maxwell} validates our predictions by directly comparing the Maxwell model to its supersymmetric Dirac theory. This procedure highlights the correspondence between traditional fermionic phases and even parity bosonic phases, while also elucidating the fundamental role of spin. Sec.~\ref{sec:Conclusions} presents our conclusions and a discussion of how to search for bosonic phases in gyrotropic plasmas and quantum wells. We anticipate the development of new experimental tools to probe the signatures of these spin-1 quantized photonic edge states. 

The focus of this paper is TRS broken topological bosonic phases which possess unidirectional edge states. As mentioned above, this is fundamentally related to optical Hall conductivity and gyrotropy in matter. However, TRS protected bosonic phases are also possible and show counter-propagating edge states \cite{van_mechelen2017}. This arises from antisymmetric magneto-electricity as opposed to gyrotropy. The hallmark of both these bosonic phases is longitudinal spin-1 quantization at the edge. These topologically protected edge states are emergent massless photons with massive-like photons in the bulk material. The rigorous validity of these topological bosonic phases follows from supersymmetric Dirac theory and constitutes a one-to-one mapping to the continuum fermionic phase. This direct analogy between Dirac-fermions and Maxwell-bosons fundamentally requires spatiotemporal dispersion, which has not been previously tackled in electromagnetic topological field theories.

\begin{figure*}
  \includegraphics[width=0.9\textwidth]{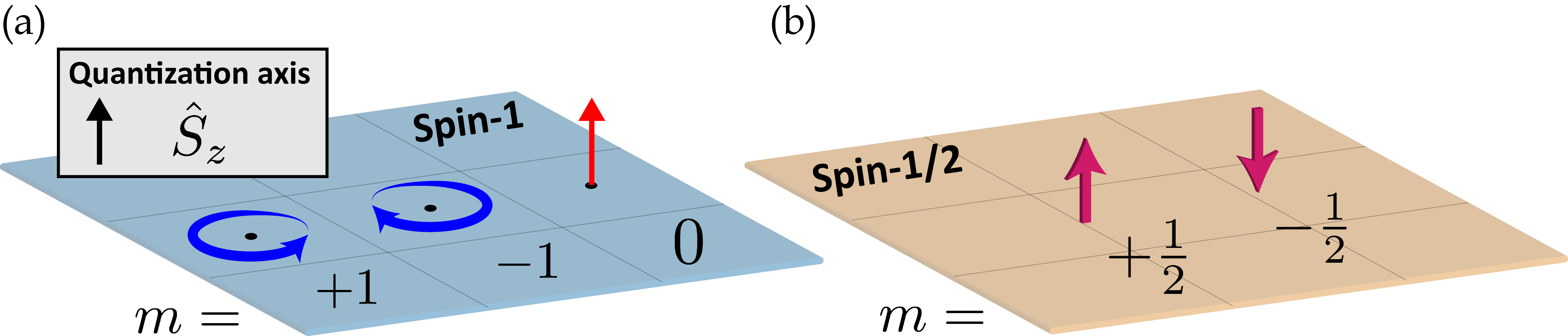}
  \caption{Our work emphasizes the fundamental differences between 2+1D topological materials for Maxwell-bosons and Dirac-fermions, which are characterized by their bulk spin quantum numbers. In 2D, the quantization axis is along $z$ as all rotations occur in the $x$-$y$ plane. Both \textbf{(a)} photonic and \textbf{(b)} electronic topologies are connected to $\hat{S}_z$ quantization at certain high-symmetry $\mathbf{k}$ points in the bulk material. The distinction lies in their rotational symmetries ($\mathcal{R}$). Photons are bosonic particles and respect spin-1 statistics $\mathcal{R}(2\pi)=+1$, which possess integer spin projections $m=\pm 1,0$. Conversely, electrons are fermionic particles and respect spin-\textonehalf{} statistics $\mathcal{R}(2\pi)=-1$, which possess half-integer spin projections $m=\pm\frac{1}{2}$. This changes the interpretation of topological invariants and the observable phenomena of different particles.}
    \label{fig:quantization}
\end{figure*}

\section{Continuum topological photonics}\label{sec:Continuum_Photonics}

\subsection{2+1D electrodynamics}

In two spatial dimensions (and one temporal dimension), the propagation of charge is restricted to the $x$-$y$ plane. This limits the degrees of freedom of both the electromagnetic field and the induced response of a material (Fig.~\ref{fig:diagram}). Therefore, we focus on strictly transverse-magnetic (TM) waves, meaning there are only 3 unique components of the field. From first principles (Appendix \ref{app:2DElectro}), we derive the corresponding wave equation of the 2D photon coupled to matter,
\begin{equation}\label{eq:wave_equation}
\mathcal{H}_0(\mathbf{k})f=\omega\mathcal{M}(\omega,\mathbf{k})f, \qquad f=\begin{bmatrix}
E_x\\ E_y \\ H_z
\end{bmatrix},
\end{equation}
where $f$ is the TM polarization state (wavefunction) of the electromagnetic field. In the absence of matter, $\mathcal{H}_0(\mathbf{k})$ are the vacuum Maxwell equations in momentum space,
\begin{equation}\label{eq:vacuum}
\mathcal{H}_0(\mathbf{k})=k_x\hat{S}_x +k_y\hat{S}_y=\begin{bmatrix}
0 & 0 & -k_y\\
0& 0 & k_x\\
-k_y &  k_x & 0 \\
\end{bmatrix}.
\end{equation}
Notice that $\mathcal{H}_0(\mathbf{k})=\mathbf{k}\cdot\mathbf{S}$ represents optical helicity, i.e. the projection of momentum $\mathbf{k}$ onto the spin $\mathbf{S}$. We identify these spin-1 operators $\hat{S}_x$ and $\hat{S}_y$ that satisfy the angular momentum algebra $[\hat{S}_x,\hat{S}_y]=i\hat{S}_z$,
\begin{equation}\label{eq:Spin}
\hat{S}_x=\begin{bmatrix}
0 & 0 & 0 \\
0 & 0 & 1 \\
0 & 1 & 0
\end{bmatrix},~~\hat{S}_y=\begin{bmatrix}
0 & 0 & -1 \\
0 & 0 & 0 \\
-1 & 0 & 0
\end{bmatrix},~~\hat{S}_z=\begin{bmatrix}
0 & -i & 0 \\
i & 0 & 0 \\
0 & 0 & 0
\end{bmatrix}.
\end{equation}
Here, $(\hat{S}_z)_{ij}=-i\epsilon_{ijz}$ is the generator of rotations in the $x$-$y$ plane and is represented by the antisymmetric matrix. $\hat{S}_z$ will be foundational when discussing spin-1 symmetries in 2D.

The linear response function of the 2D material $\mathcal{M}$ is dependent on continuous variables $\omega$ and $\mathbf{k}$,
\begin{equation}
\mathcal{M}(\omega,\mathbf{k})=\begin{bmatrix}
\varepsilon_{xx} & \varepsilon_{xy} & \chi_{x} \\
\varepsilon_{xy}^* & \varepsilon_{yy} & \chi_{y}\\
\chi_{x}^*& \chi_{y}^* & \mu
\end{bmatrix}, ~~ \begin{array}{ll}
D_i=\varepsilon_{ij}E^j+\chi_{i} H_z, \\
~\\
B_z=\chi_{i}^*E^i+\mu H_z,\\
\end{array}
\end{equation}
which compactly represents the constitutive relations. We include all possible material responses as a generalization - for instance magneto-electricity $\chi_i$ and birefringence in $\varepsilon_{ij}$. However, based on symmetry constraints, we will show that only certain parameters of $\mathcal{M}$ are important in the topological classification.

\subsection{Continuum response function}

Alas, Eq.~(\ref{eq:wave_equation}) poses a problem; it does not represent a proper first-order in time Hamiltonian since the response function $\mathcal{M}(\omega,\mathbf{k})$ is dependent on its own eigenvalue. Nevertheless, we can prove that it is derived from a first-order Hamiltonian by exploiting stringent symmetry properties. We demand Hermiticity $\mathcal{M}=\mathcal{M}^\dagger$ such that the response is lossless. We also require positive definiteness $\bar{\mathcal{M}}=\partial_\omega(\omega\mathcal{M})>0$ to ensure the energy density is non-negative and admits proper normalization $f^\dagger\bar{\mathcal{M}}f>0$. The response must be causal (Kramers-Kronig) and obey the reality condition $\mathcal{M}(\omega,\mathbf{k})=\mathcal{M}^*(-\omega,-\mathbf{k})$, guaranteeing the electromagnetic fields are real-valued \cite{landau2013electrodynamics}. Two additional constraints should also be considered for realistic materials. Stability at static equilibrium $\mathcal{M}(0,\mathbf{k})=\bar{\mathcal{M}}(0,\mathbf{k})>0$, and the ultraviolet limit $\lim_{\omega\to\infty}\mathcal{M}(\omega,\mathbf{k})=\mathds{1}_3$. Here, $\mathds{1}_3$ is the $3\times 3$ identity matrix and the limit implies transparency at high-frequency $\omega\to\infty$, as the material cannot respond to sufficiently fast temporal oscillations.

Combining all the above criteria, we find that the response function can always be decomposed as a discrete summation of oscillators \cite{Haldane2008,Silveirinha2015,Philbin2014},
\begin{equation}\label{eq:Response}
\mathcal{M}(\omega,\mathbf{k})=\mathds{1}_3-\sum_\alpha \frac{\mathcal{C}^\dagger_{\alpha \mathbf{k}} \mathcal{C}_{\alpha\mathbf{k}}}{\omega_{\alpha \mathbf{k}}(\omega-\omega_{\alpha \mathbf{k}})}.
\end{equation}
$\alpha$ labels any arbitrary bosonic excitation in the material, such as an exciton or phonon, which couples linearly to the electromagnetic fields via the $3\times 3$ tensor $\mathcal{C}_{\alpha\mathbf{k}}$. $\omega_{\alpha\mathbf{k}}$ is the resonant energy of the oscillator and corresponds to a first-order pole of the response function. Note both $\mathcal{C}_{\alpha\mathbf{k}}$ and $\omega_{\alpha\mathbf{k}}$ are in general $\mathbf{k}$ dependent. We emphasize that the response function is consistent with previous work on gyrotropic plasmas \cite{Nikolaev1996,eroglu2010wave}. However, our key advance is that the tensors $\mathcal{C}_{\alpha\mathbf{k}}$, characterizing the collective light-matter excitations, carry information of spin and orbital angular momentum.

\subsection{Continuum Hamiltonian}

A detailed derivation of the continuum electromagnetic Hamiltonian $H(\mathbf{k})$ is presented in Appendix \ref{app:Oscillators}. To accomplish this, we expand the response function $\mathcal{M}(\omega,\mathbf{k})$ in terms of 3-component matter oscillators $\psi_\alpha$. These represent internal polarization and magnetization modes of the material, 
\begin{equation}\label{eq:oscillator}
\psi_\alpha=\frac{\mathcal{C}_{\alpha\mathbf{k}}f}{\omega-\omega_{\alpha\mathbf{k}}}, \qquad \omega \psi_\alpha=\omega_{\alpha\mathbf{k}}\psi_\alpha+\mathcal{C}_{\alpha\mathbf{k}}f.
\end{equation}
We now define $u$ as the generalized state vector of the electromagnetic problem, accounting for the photon $f$ and all possible internal excitations $\psi_\alpha$,
\begin{equation}
H(\mathbf{k})u=\omega u, \qquad
u=\begin{bmatrix}
f & \psi_1 & \psi_2 & \ldots
\end{bmatrix}^\intercal,
\end{equation}
which satisfies a first-order Hamiltonian wave equation. Notice that contraction of $u$ naturally reproduces the energy density upon summation over all degrees of freedom
$u^\dagger u=f^\dagger \bar{\mathcal{M}}f$, with $ \bar{\mathcal{M}}=\partial_\omega(\omega\mathcal{M})>0$ always positive definite. The continuum Hamiltonian $H(\mathbf{k})$ acting on $u$ is given concisely as,
\begin{equation}\label{eq:Hamiltonian}
H(\mathbf{k})=\begin{bmatrix}
\mathcal{H}_0(\mathbf{k})+\sum_\alpha \omega_{\alpha\mathbf{k}} ^{-1}\mathcal{C}^\dagger_{\alpha\mathbf{k}} \mathcal{C}_{\alpha\mathbf{k}} & ~~\mathcal{C}^\dagger_{1\mathbf{k}}~~ & ~~\mathcal{C}^\dagger_{2\mathbf{k}}~~ & \ldots \\
\mathcal{C}_{1\mathbf{k}} & \omega_{1\mathbf{k}} & 0 & \ldots\\
\mathcal{C}_{2\mathbf{k}} & 0 & \omega_{2\mathbf{k}} & \ldots\\
\vdots & \vdots & \vdots & \ddots 
\end{bmatrix}.
\end{equation}
This eigenvalue problem generates the complete spectrum of quasiparticle eigenstates,
\begin{equation}\label{eq:Eigenstates}
H_\mathbf{k}u_{n\mathbf{k}}=\omega_{n\mathbf{k}}u_{n\mathbf{k}},
\end{equation}
and the eigenstates are normalized to the energy density $u^\dagger_{n\mathbf{k}}u_{n\mathbf{k}}=f^\dagger_{n\mathbf{k}}\bar{\mathcal{M}}(\omega_{n\mathbf{k}},\mathbf{k})f_{n\mathbf{k}}=1$. Moreover, the eigenenergies $\omega_{n\mathbf{k}}$ are the $n$ non-trivial roots of the characteristic equation,
\begin{equation}\label{eq:characteristic}
\det[\mathcal{H}_0(\mathbf{k})-\omega\mathcal{M}(\omega,\mathbf{k})]=0, \qquad \omega=\omega_n(\mathbf{k}),
\end{equation}
proving that the response function $\mathcal{M}(\omega,\mathbf{k})$ is derived from a first-order Hamiltonian $H(\mathbf{k})$.

\subsection{Continuum regularization (one-point compactification)}

Our goal is to develop a continuum topological theory that accounts for both spatiotemporal dispersion and the inherent bosonic properties of light. Due to the unbounded nature of the momentum space $\mathbb{R}^2$, continuum Chern numbers are usually ill-defined. Nevertheless, as long as the system is properly \textbf{regularized}, continuum field theories are possible and can be incredibly powerful tools to study long wavelength topological physics \cite{Shen2010,Shenoy2012,SHEN2011}. A necessary condition is one-point compactification of the momentum space \cite{Ryu2010,munkres2000topology,bernevig2013topological}, which governs the high-$k$ asymptotic behavior of the Hamiltonian. This requirement is well understood in condensed matter and demands the Hamiltonian approach a \textbf{directionally independent} value,
\begin{equation}\label{eq:onepoint}
\lim_{k\to \infty}H(\mathbf{k})\to H(k),
\end{equation}
where $k=\sqrt{\mathbf{k}\cdot\mathbf{k}}$ is the magnitude of the wavevector. In this way, all limits at infinity are mapped into the same point and satisfy a ``periodic'' boundary condition. The momentum space is closed and topologically equivalent to the Reimann sphere $\mathbb{R}^2\simeq S^2$ (Fig.~\ref{fig:Riemann_Sphere}). Hence, Chern numbers are quantized. A rigorous proof is presented in Appendix \ref{app:Continuum}.

This constraint has important implications in continuum photonic media. Since Maxwell's equations are strictly first-order in spatial derivatives (Eq.~(\ref{eq:vacuum})), one-point compactification can only be satisfied by introducing non-locality \cite{agranovich2013crystal,Silin2961}. Non-locality (or spatial dispersion) is the momentum dependence of linear response - commonly ignored in photonics problems, DC transport measurements, as well as Faraday rotation experiments. However, we strongly emphasize that the high-$k$ behavior cannot be neglected even in the long wavelength continuum theory. These deep sub-wavelength components encode global information of the fields and are essential to properly describe the topological physics. By exploiting rotational symmetry, we will show that the asymptotic behavior of the Hamiltonian $H(\mathbf{k})$ and by extension, the response function $\mathcal{M}(\omega,\mathbf{k})$, is naturally regularized and predicts new bosonic phases of matter.

\begin{figure}
\includegraphics[width=0.8\columnwidth]{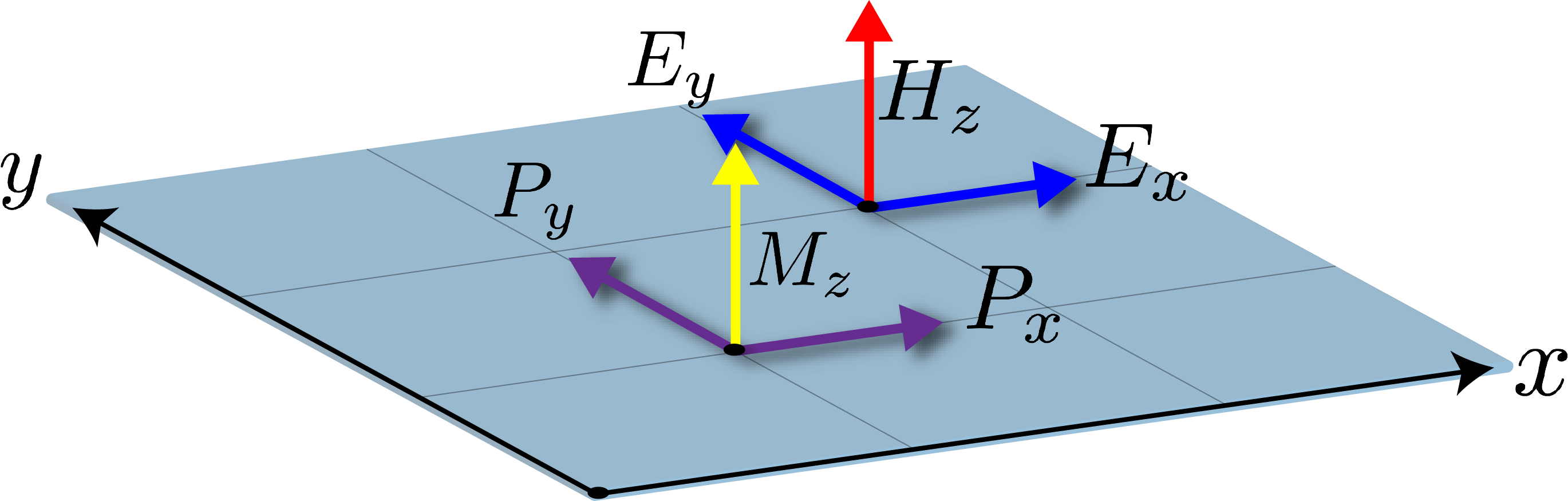}
\caption{Only transverse-magnetic (TM) waves propagate as charge is restricted to the $x$-$y$ plane (blue and red arrows denote the fields).  This limits the degrees of freedom of both the electromagnetic field and the induced response of a material. Electromagnetic polarization and magnetization response in a 2D material is shown with the purple and yellow arrows. The electric and magnetic displacement fields are the linear superposition of $D_i=P_i+E_i$ and $B_z=M_z+H_z$. Our focus in this paper is gyrotropic media which correspond to optical (dynamical) Hall conductivity.}
\label{fig:diagram}
\includegraphics[width=0.8\linewidth]{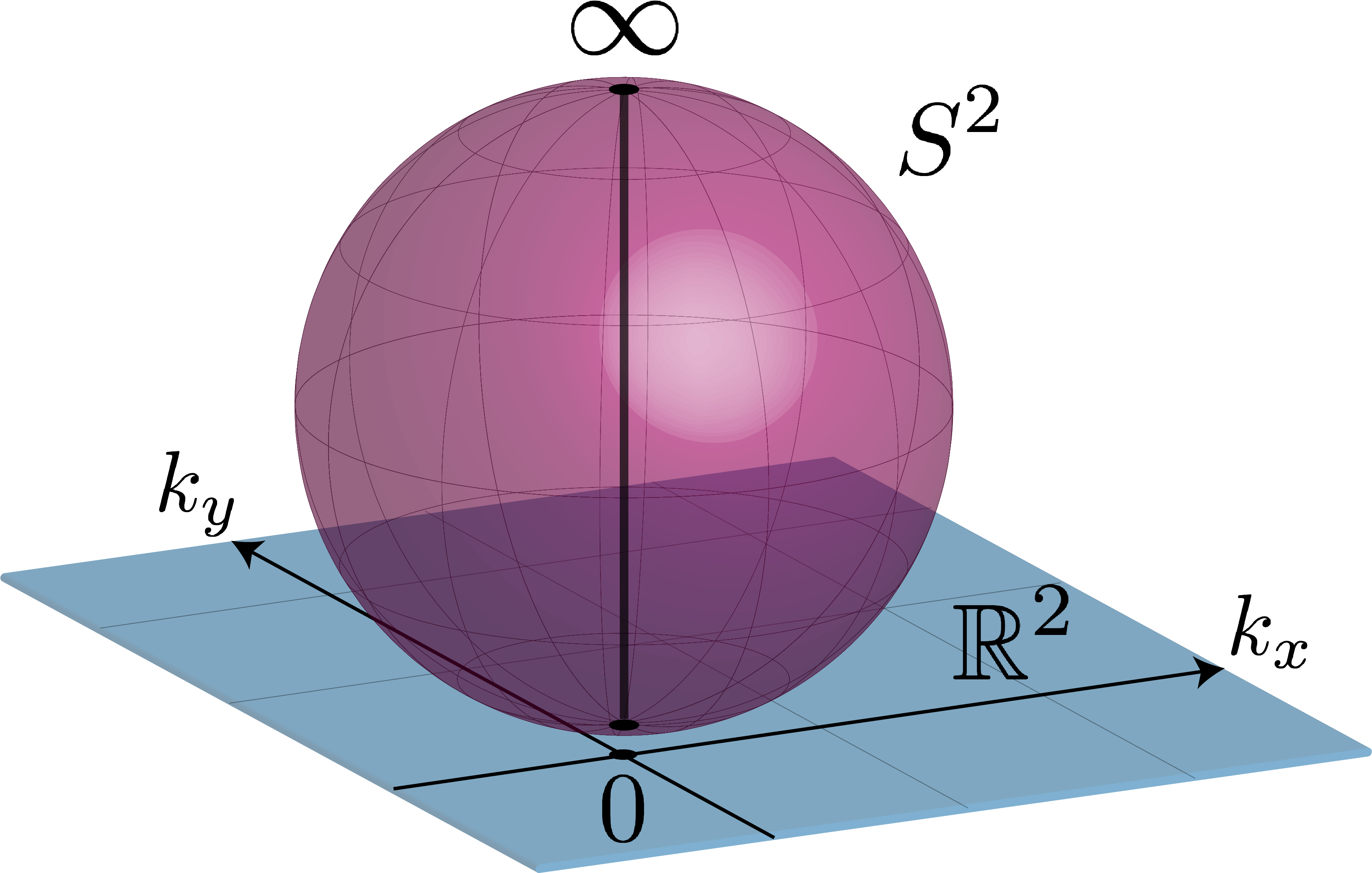}
\caption{One-point compactification of the momentum space $\mathbb{R}^2\simeq S^2$ over which the topological quantum numbers are defined. When the Hamiltonian is properly regularized, the planar $\mathbf{k}$-space is topologically equivalent to the bounded Reimann sphere. $k_p=0$ and $k_p=\infty$ are the rotationally invariant (high-symmetry) points on the sphere, passing through the $z$-axis. This procedure is necessary to ensure Chern quantization in continuum topological field theories and fundamentally requires non-local photonic media.}
\label{fig:Riemann_Sphere}
\end{figure}

\section{Rotational symmetry}\label{sec:Rotational_Symmetry}

\subsection{Definition of orbital, spin and total angular momentum}

If the two dimensional crystal has a center (at least 3-fold cyclic \cite{Hestenes2002,nye1985physical}), the continuum Hamiltonian is rotationally symmetric about $z$,
\begin{equation}\label{eq:rotational}
\mathcal{R}^{-1}H(\mathcal{R}\mathbf{k})\mathcal{R}=H(\mathbf{k}), \qquad \mathcal{R}(2\pi)=\mathds{1}_3,
\end{equation}
and the eigenenergies $\omega=\omega_n(k)$ depend only on the magnitude of $k$. Note that $\mathcal{R}$ is diagonal in $u$, meaning the photon and each oscillator is rotated individually, $f\to \mathcal{R}f$ and $\psi_\alpha\to \mathcal{R}\psi_\alpha$. In this case $\mathcal{R}(\theta)$ is a continuous rotation,
\begin{equation}
\mathcal{R}(\theta)=\exp[i\theta\hat{S}_z]=\begin{bmatrix}
\cos\theta & \sin\theta & 0 \\
-\sin\theta & \cos\theta & 0 \\
0 & 0 &1
\end{bmatrix},
\end{equation}
and can be expressed as the exponential of the spin-1 generator $(\hat{S}_z)_{ij}=-i\epsilon_{ijz}$. This represents an element of SO(3) in the subspace of $\mathbb{R}^2$ \cite{hall_lie_2015}, as all rotations occur in the $x$-$y$ plane. We stress that the vector representation is \textbf{bosonic}, meaning the quasiparticles return in phase under cyclic revolution $\mathcal{R}(2\pi)=\mathds{1}_3$.

Since the Hamiltonian possesses a \textbf{continuous} rotational symmetry, the total angular momentum (TAM) is conserved,
\begin{equation}\label{eq:TAM}
[\hat{J}_z,H(\mathbf{k})]=0, \qquad \hat{J}_z=\hat{L}_z+\hat{S}_z.
\end{equation}
Eq.~(\ref{eq:rotational}) and (\ref{eq:TAM}) are equivalent statements in this context. Here, $\hat{L}_z$ is the orbital angular momentum (OAM) operator in 2D $\mathbf{k}$-space and can be expressed in polar coordinates as,
\begin{equation}
\hat{L}_z=-ik_x\frac{\partial}{\partial k_y}+ik_y\frac{\partial}{\partial k_x}=-i\partial_\phi.
\end{equation}
Eigenstates of the OAM are well known and represent quantized azimuthal charges,
\begin{equation}
\hat{L}_z |l\rangle=l|l\rangle, \qquad |l\rangle=\exp(il\phi),
\end{equation}
where $l\in \mathbb{Z}$ is any integer.

Conversely, eigenstates of the spin angular momentum (SAM) represent states of quantized polarization, transverse to the $x$-$y$ plane,
\begin{equation}\label{eq:SAM_eigenstate}
\hat{S}_z\mathbf{e}=m\mathbf{e}.  
\end{equation}
The matrix form of $\hat{S}_z$ is given in Eq.~(\ref{eq:Spin}). For photons, the spin is an integer $m=\pm 1,0$ and takes one of three discrete values. First, we have the $m=\pm 1$ spin states,
\begin{equation}
\mathbf{e}_\pm=\frac{1}{\sqrt{2}}\begin{bmatrix}
1\\ \pm i \\0
\end{bmatrix}, \qquad \hat{S}_z\mathbf{e}_\pm=\pm\mathbf{e}_\pm.
\end{equation}
$\mathbf{e}_\pm$ are resonant electric ($H_z=0$) counter-rotating states. Secondly, we have the $m=0$ spin state, which is resonant magnetic ($E_i=0$) and irrotational, 
\begin{equation}
 \mathbf{e}_0=\begin{bmatrix}
0\\ 0 \\ 1
\end{bmatrix}, \qquad \hat{S}_z\mathbf{e}_0=0.
\end{equation}
A visualization of the quantized spin-1 states is displayed in Fig.~\ref{fig:quantization}(a) and this is compared to quantized spin-\textonehalf{} states in Fig.~\ref{fig:quantization}(b). In Sec.~\ref{sec:Bosonic_Phases}, we will prove that these spin quantized eigenstates naturally arise at high-symmetry $\mathbf{k}$ points in 2+1D bosonic phases.

\subsection{High-symmetry points and gauge singularities}

At an arbitrary momentum $\mathbf{k}$, the quasiparticles $u_{n\mathbf{k}}$ are not eigenstates of $\hat{L}_z$ or $\hat{S}_z$. Instead, they are eigenstates of the total angular momentum $\hat{J}_z=\hat{L}_z+\hat{S}_z$,
\begin{equation}\label{eq:TAMgauge}
\hat{J}_z u_{n\mathbf{k}}=j_nu_{n\mathbf{k}}, \qquad j_n \in \mathbb{Z},
\end{equation}
where $j_n$ is an integer for bosons. Since $\hat{J}_z$ is a differential operator, the choice of $j_n$ represents a particular Berry gauge for the eigenstates. This gauge is single-valued for all $\mathbf{k}$ with the possible exception of two points, $k_p=0$ and $k_p=\infty$. These are called \textbf{high-symmetry points} (HSPs). At these specific momenta, the Hamiltonian is \textbf{rotationally invariant} \cite{Fang2012},
\begin{equation}\label{eq:HSPs}
\mathcal{R}^{-1}H(k_p)\mathcal{R}=H(k_p), \qquad [\hat{S}_z,H(k_p)]=0,
\end{equation}
which follows immediately from Eq.~(\ref{eq:rotational}) and (\ref{eq:TAM}). In the continuum theory, $k_p=0$ is a HSP because the origin always rotates into itself. Owing to one-point compactification (Eq.~(\ref{eq:onepoint})), $k_p=\infty$ is also a HSP. This is clear by direct inspection of the Riemann sphere in Fig.~\ref{fig:Riemann_Sphere}. A rotation in the plane of $\mathbb{R}^2$ rotates $S^2$ about its axis, keeping both $k_p=0$ and $k_p=\infty$ fixed. Invariance at $k_p=\infty$ is therefore imperative to describe continuum topological theories.

At HSPs the SAM of any eigenstate $u_{n\mathbf{k}}$ is quantized and this is guaranteed by symmetry (Eq.~(\ref{eq:HSPs})). Still, the Berry gauge may be multi-valued here due to the OAM - known as a phase singularity \cite{Heckenberg1992},
\begin{equation}\label{eq:gauge}
\lim_{k\to k_p}u_{n}(\mathbf{k})\to u_{n}(k_p)\exp\left[il_n(k_p)\phi\right],
\end{equation}
\begin{equation}
\hat{S}_z u_{n}(k_p)=m_n(k_p)u_{n}(k_p),
\end{equation}
where $j_n=l_n(k_p)+m_n(k_p)$ at HSPs. We come to an important revelation from Eq.~(\ref{eq:gauge}). If the spin \textbf{does not change} within the eigenstate dispersion $m_n(0)=m_n(\infty)$, we can remove the phase singularity at both points simultaneously $l_n(0)=l_n(\infty)=0$, such that the Berry gauge $j_n=m_n(0)=m_n(\infty)$ is single-valued for all $\mathbf{k}$.

However, if the spin \textbf{changes} within the dispersion $m_n(0) \neq m_n(\infty)$, this procedure is impossible. The Berry gauge is always multi-valued because the singularity $l_n(k_p)\neq 0$ cannot be resolved at $k_p=0$ and $k_p=\infty$ simultaneously. This is a non-trivial topology. The physical interpretation is simple but profound; since the TAM is conserved for each eigenstate $\Delta j_n=0$, the OAM $\Delta l_n= l_n(\infty)-l_n(0)\neq 0$ must change to compensate for the SAM,
\begin{equation}\label{eq:Quantum_Jump}
\Delta l_n=-\Delta m_n= m_n(0)-m_n(\infty).
\end{equation}
We will now prove that Eq.~(\ref{eq:Quantum_Jump}) fundamentally defines the Chern classification of 2+1D bosonic phases.

\section{Continuum topological bosonic phases}\label{sec:Bosonic_Phases}

\subsection{Continuum photonic Chern number}

Utilizing the eigenstates of the Hamiltonian in Eq.~(\ref{eq:Eigenstates}); we obtain the Berry connection by varying a quasiparticle with respect to the momentum,
\begin{equation}
\mathbf{A}_n(\mathbf{k})=-iu_{n\mathbf{k}}^\dagger\partial_\mathbf{k} u_{n\mathbf{k}}.
\end{equation}
Applying the curl produces the gauge invariant Berry curvature $F_n(\mathbf{k})=\hat{\mathbf{z}}\cdot[\partial_\mathbf{k}\times\mathbf{A}_n(\mathbf{k})]$. The Chern number $C_n$ is a global topological invariant and is traditionally found by integrating $F_n$ over the 2D Brillouin zone - i.e. the torus $\mathbb{T}^2=S^1\times S^1$. For continuum theories, we integrate over the entire 2D momentum space $\mathbb{R}^2$,
\begin{equation}
C_n=\frac{1}{2\pi}\iint_{\mathbb{R}^2}F_n(\mathbf{k}) ~d^2\mathbf{k}.
\end{equation}
When properly regularized, the planar manifold is topologically equivalent to the Riemann sphere $\mathbb{R}^2\simeq S^2$ and the Chern number is quantized (Appendix \ref{app:Continuum}).

Although photonic Chern numbers have been defined, neither the high-$k$ behavior nor the inherent bosonic properties have been addressed. With this in mind, we return to the Berry connection $\mathbf{A}_n$ in polar coordinates $\partial_\mathbf{k}=\hat{\mathbf{k}}\partial_k +\hat{\pmb{\phi}}\partial_\phi$,
\begin{equation}
\mathbf{A}_n(\mathbf{k})=\hat{\mathbf{k}}A_n^k(k)+\hat{\pmb{\phi}}A_n^\phi(k).
\end{equation}
Due to rotational symmetry, the polar components of $\mathbf{A}_n$ depend only on $k$. Furthermore, we can connect the Berry potential $A_n^\phi$ directly to the OAM,
\begin{equation}
A_n^\phi(k)= -iu_{n\mathbf{k}}^\dagger\partial_\phi u_{n\mathbf{k}}=\langle \hat{L}_z\rangle_n.
\end{equation}
Here, $\langle\hat{L}_z\rangle_n$ is the expectation value of the OAM for the $n^\textrm{th}$ eigenstate. This corresponds to a Berry curvature $F_n$ of,
\begin{equation}
F_n(k)=\partial_k \langle \hat{L}_z\rangle_n.
\end{equation}
When integrating over all momenta $d^2\mathbf{k}=dkd\phi$, we find that the continuum Chern number $C_n$ is determined solely by the phase singularities at HSPs,
\begin{equation}
C_n=\int_0^\infty dk ~\partial_k \langle \hat{L}_z\rangle_n=\langle \hat{L}_z\rangle_n|_0^\infty=\Delta l_n,
\end{equation}
precisely the change in OAM. Substituting for $\Delta l_n=-\Delta m_n$ in Eq.~(\ref{eq:Quantum_Jump}), we attain an elegant expression for the Chern number,
\begin{equation}\label{eq:Chern}
C_n=\Delta l_n=m_n(0)-m_n(\infty).
\end{equation}
Eq.~(\ref{eq:Chern}) is one of the central results of this paper and is valid for both fermionic and bosonic representations. Essential differences between the two are immediately apparent.

For a spin-\textonehalf{} electron, quanta take one of two half-integer values $m_n=\pm \frac{1}{2}$. Consequently, we find only one truly distinct fermionic phase,
\begin{equation}
\textrm{fermion}: \qquad C_n=\pm 1, 0.
\end{equation}
However, for the spin-1 photon, quanta take three integer values $m_n=\pm 1, 0$. We discover two unique bosonic phases,
\begin{equation}
\textrm{boson}: \qquad C_n=\pm 2, \pm 1, 0.
\end{equation}
One with even parity $C_n=\pm 2$ and one with odd $C_n=\pm 1$. Even parity corresponds to a change from $m_n(0)=\pm 1$ to $m_n(\infty)=\mp 1$ at HSPs. This phase is familiar in interacting bosonic systems and is identified with a \textbf{single} gapless boson at the edge \cite{Chen1604,Metlitski2013,Vishwanath2013,Senthil2013,Vishwanath2012,Wen2011} - not two as we might expect from fermionic Chern number arguments. Odd parity bosonic phases are quite exotic in this regard \cite{Sheng2016,Read1998}. This phase corresponds to a change from $m_n(0)=\{0,\pm 1\}$ to $m_n(\infty)=\{\pm 1,0\}$ at HSPs.

\subsection{Non-local regularization of the response function}

We now derive the asymptotic behavior of the Hamiltonian $H(\mathbf{k})$ to ensure the continuum theory is properly regularized at $k\to\infty$. This will help us discover the precise form of the response function $\mathcal{M}(\omega,\mathbf{k})$ and the order of non-locality necessary to describe a topological field theory. Non-locality plays two equally important roles in this context - it distinguishes between \textbf{trivial} and \textbf{non-trivial} phases. If high-$k$ components are ignored, it is impossible to define either of these phases in the continuum.

From Eq.~(\ref{eq:Response}) and (\ref{eq:Hamiltonian}), rotational symmetry implies the coupling tensors obey $\mathcal{R}^{-1}\mathcal{C}_\alpha(\mathcal{R}\mathbf{k})\mathcal{R}=\mathcal{C}_\alpha(\mathbf{k})$ and the oscillator resonances $\omega_\alpha(k)$ depend only on $k$. We find the exact expression of $\mathcal{C}_\alpha(\mathbf{k})$,
\begin{equation}
\mathcal{C}_\alpha(\mathbf{k})=c_\alpha(k)\mathbf{k}\otimes\mathbf{k}+d_\alpha(k)\mathbf{k}\cdot\mathbf{S}+\mathcal{G}_{\alpha}(k),
\end{equation}
where $c_\alpha(k)$ and $d_\alpha(k)$ are scalars. It is easy to check that the tensors also commute with $[\hat{J}_z,\mathcal{C}_\alpha(\mathbf{k})]=0$, conserving TAM. $c_{\alpha}(k)$ introduces a non-local birefringence in $\varepsilon_{ij}$ and $d_{\alpha}(k)$ is a type of non-local magneto-electricity $\chi_i$. Both terms are permitted by symmetry but neither is important, as all contributions besides $\mathcal{G}_{\alpha}(k)$ vanish identically at $k_p=0$ and $k_p=\infty$. This is because $\mathcal{G}_{\alpha}(k)$ is the only rotationally invariant component of $\mathcal{C}_\alpha(\mathbf{k})$, which defines the topology,
\begin{equation}\label{eq:G_matrix}
\mathcal{R}^{-1}\mathcal{G}_{\alpha}(k)\mathcal{R}=\mathcal{G}_{\alpha}(k), \qquad [\hat{S}_z,\mathcal{G}_{\alpha}(k)]=0.
\end{equation}
The Hamiltonian in Eq.~(\ref{eq:Hamiltonian}) takes the following form at HSPs,
\begin{equation}\label{eq:reg_Hamiltonian}
H(k_p)=\begin{bmatrix}
\sum_\alpha \omega_{\alpha k_p} ^{-1}\mathcal{G}^\dagger_{\alpha k_p} \mathcal{G}_{\alpha k_p} & ~~\mathcal{G}^\dagger_{1 k_p}~~ & ~~\mathcal{G}^\dagger_{2 k_p}~~ & \ldots \\
\mathcal{G}_{1 k_p} & \omega_{1 k_p} & 0 & \ldots\\
\mathcal{G}_{2 k_p} & 0 & \omega_{2 k_p} & \ldots\\
\vdots & \vdots & \vdots & \ddots 
\end{bmatrix}.
\end{equation}
Notice the vacuum Maxwell equations $\mathcal{H}_0(\mathbf{k})$ play no role in either limit; the Hamiltonian is governed entirely by the material response at HSPs. Nevertheless, this imposes pivotal stipulations on the asymptotic behavior. The largest powers in $k$ must arise from $\mathcal{G}_{\alpha k}$ as these terms dominate at exceedingly large momentum $k\to\infty$. Consequently, $\mathcal{G}_{\alpha k}$ and $\omega_{\alpha k}$ require quadratic non-locality $\propto k^2$ \textbf{at minimum}, since the vacuum fields $\mathcal{H}_0(\mathbf{k})$, which are linear in $\mathbf{k}$, must be outpaced in the $k\to \infty$ limit.

By extension of Eq.~(\ref{eq:reg_Hamiltonian}), the response function is regularized and rotationally invariant at HSPs,
\begin{equation}\label{eq:response_commute}
[\hat{S}_z,\mathcal{M}(\omega,k_p)]=0.
\end{equation}
Upon summation over all oscillators describing the linear response, $\mathcal{M}$ takes a remarkably simple form,
\begin{equation}\label{eq:response_HSPs}
\begin{split}
\mathcal{M}(\omega,k_p)= \mathds{1}_3-\sum_\alpha \frac{\mathcal{G}^\dagger_{\alpha k_p} \mathcal{G}_{\alpha k_p}}{\omega_{\alpha k_p}\left(\omega-\omega_{\alpha k_p}\right)}
=\begin{bmatrix}
\varepsilon & ig & 0\\
-ig & \varepsilon & 0\\
0 & 0& \mu
\end{bmatrix},
\end{split}
\end{equation}
where all parameters are evaluated at $k_p$. Here, $\varepsilon$ and $\mu$ are the conventional scalar permittivity and permeability of a 2D material. $g$ is a generalized gyrotropic coupling which breaks both parity and time-reversal symmetry.

Although the condition at $k_p=\infty$ is a mathematical requisite, it makes perfect sense physically when we acknowledge that the continuum theory is simply an approximation of the underlying crystal lattice. In reality, the momentum can never reach arbitrarily large values. As the momentum approaches the scale of the lattice constant $ka\approx \pi $, the wave approaches a Bragg condition. These are HSPs in the reciprocal lattice \cite{Fang2012,Hestenes2002,hall_lie_2015,nye1985physical} so the continuum theory must encode this behavior. Accordingly, the $k\to \infty$ limit should be interpreted as a Bragg resonance.

\subsection{Transverse spin quantization of the photon}

We go one step further to uncover the precise origin of the spin-1 eigenvalues $m_n(k_p)$, the spin states $\mathbf{e}$, and their relation to the response function $\mathcal{M}$. At HSPs, the SAM expectation value is represented as,
\begin{equation}\label{eq:SAM_HSPs}
\langle \hat{S}_z\rangle_n=m_n(k_p)=u^\dagger_n(k_p)\hat{S}_zu_n(k_p).
\end{equation}
Using Eq.~(\ref{eq:oscillator}) and (\ref{eq:G_matrix}), this can be simplified to yield,
\begin{equation}\label{eq:SAM_photon}
m_n(k_p)=f_n^\dagger(k_p)\bar{\mathcal{M}}(\omega_n(k_p),k_p)\hat{S}_z f_n(k_p).
\end{equation}
We note that precisely at HSPs, the quantum of spin $m_n(k_p)$ is determined entirely by the photonic component $f_n(k_p)$ of the eigenmode $u_n(k_p)$ - but not the coordinates of the matter oscillations $\psi_{\alpha}$. Utilizing the normalization condition $f^\dagger_{n}(k_p)\bar{\mathcal{M}}(\omega_n(k_p),k_p)f_n(k_p)=1$, Eq.~(\ref{eq:SAM_photon}) leads to,
\begin{equation}
\hat{S}_z f_n(k_p)=m_n(k_p)f_n(k_p).
\end{equation}
This indicates that the electromagnetic wavefunction $f$ must be a spin state $f_n(k_p)\propto \mathbf{e}$ at HSPs (Eq.~(\ref{eq:SAM_eigenstate})).

Our problem reduces to finding the eigenstates of the photon at HSPs and directly evaluating their spin eigenvalues. We return to the characteristic equation in Eq.~(\ref{eq:characteristic}), which defines the photonic wavefunction $f$. As $k\to 0$, the vacuum equations vanish identically $\mathcal{H}_0(\mathbf{k})\to 0$. Moreover, the response function is regularized and includes quadratic non-locality $\propto k^2$ at minimum. As $k\to\infty$, the vacuum fields do not contribute $\mathcal{H}_0(\mathbf{k})\to 0$. Therefore, a non-trivial solution exists $\omega_n(k_p)\neq 0$ if and only if it satisfies, 
\begin{equation}\label{eq:Bragg}
\det[\mathcal{M}(\omega_{n}(k_p),k_p)]=0.
\end{equation}
Eq.~(\ref{eq:Bragg}) represents the threshold condition at $k_p=0$ and the Bragg condition at $k_p=\infty$ for any particular eigenstate $n$. To allow for non-trivial solutions $\mathcal{M}f=0$ in Eq.~(\ref{eq:response_HSPs}), one of three possible conditions must be fulfilled,
\begin{equation}
\frac{g(\omega_n(k_p),k_p)}{\varepsilon(\omega_n(k_p),k_p)}=\pm 1, ~~~~\textrm{or} ~~~~ \mu(\omega_n(k_p),k_p)=0.
\end{equation}
We see that the photonic wavefunction is clearly a spin-1 eigenstate $f_n(k_p)\propto \mathbf{e}$ at HSPs. The gyrotropic constraint gives us counter-rotating spin states $\hat{S}_z \mathbf{e}_\pm=\pm \mathbf{e}_\pm$ with eigenvalues $m_n(k_p)=\pm 1$, while the magnetic constraint gives us the irrotational spin state $\hat{S}_z \mathbf{e}_0=0$ with eigenvalue $m_n(k_p)=0$. Physically, these conditions at HSPs correspond to gyrotropic or magnetic plasmon resonances in the bulk 2D material. In a lattice theory, the resonance at $k_p=0$ describes the response at the $\Gamma$ point, while $k_p=\infty$ describes the behavior near the edges of the Brillouin zone.

The meaning behind each topological bosonic phase is now revealed. In the even parity phase $C_n=\pm 2$, a gyrotropic mode dominates but the handedness of the plasmon changes at HSPs, $g/\varepsilon=\pm 1\to \mp 1$. If the handedness does not change as $k\to\infty$, the phase is trivial $C_n=0$. The odd parity phase $C_n=\pm 1$ is very different however. Instead, the mode changes from a magnetic plasmon $\mu=0$ to a gyrotropic plasmon $g/\varepsilon=\pm 1$ at HSPs. In the following sections, we restrict our discussion to the even parity phase $C_n=\pm 2$. The odd parity phase $C_n =\pm 1$ is significantly more complicated and will be dedicated to a future paper.

\begin{figure}
\includegraphics[width=\columnwidth]{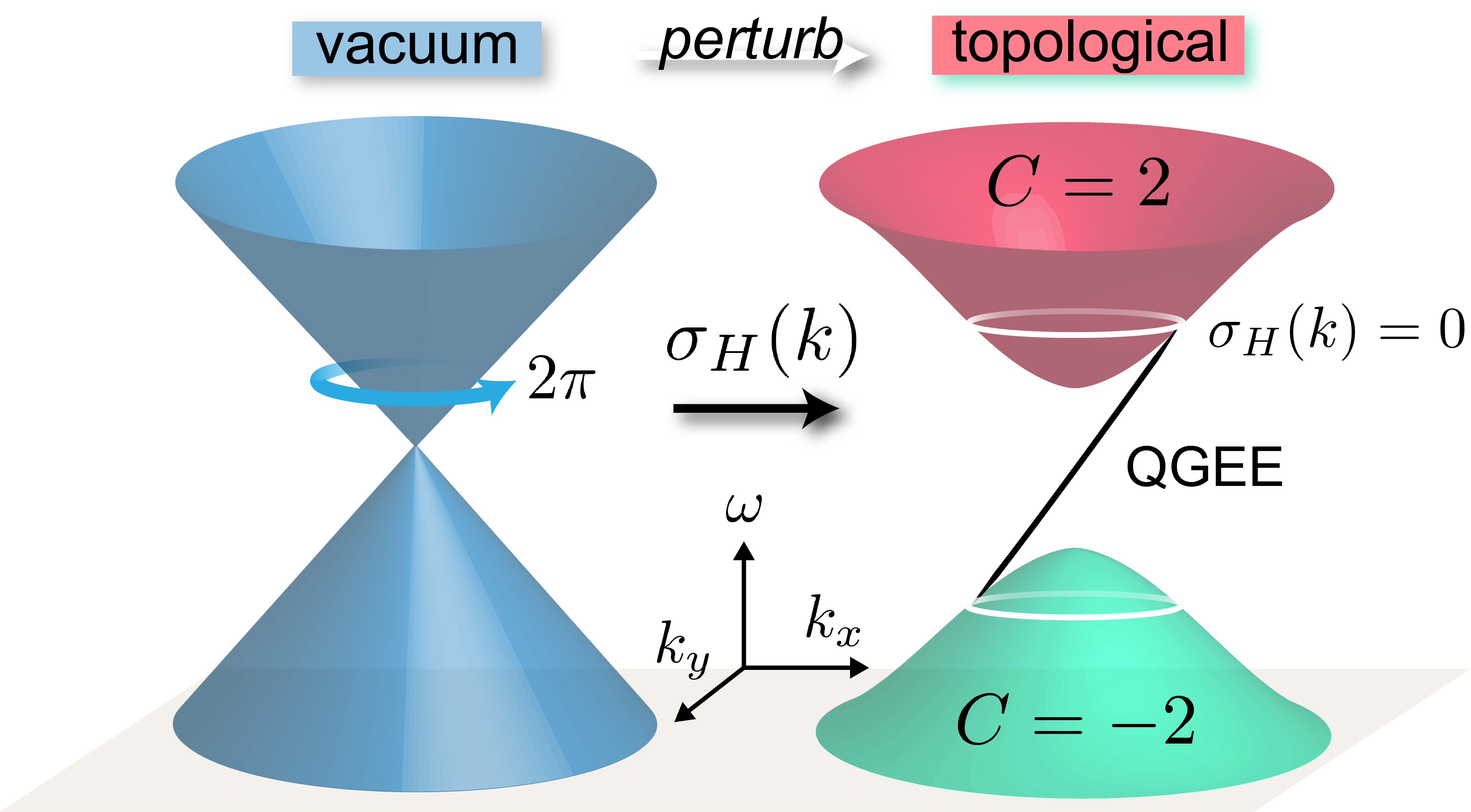}
\caption{Schematic of the exactly solvable topological model. In vacuum, Maxwell's equations can be written in the form $\mathcal{H}_0(\mathbf{k})=\mathbf{k}\cdot\mathbf{S}$, which captures both the spin-1 behavior and linear dispersion of the massless photon. The gyrotropic medium perturbs the linear dispersion and induces a bulk bandgap near zero frequency. In this case, the perturbation is a non-local Hall conductivity $\sigma_H(k)=\sigma_0-\sigma_2(ka)^2$, which behaves identically to the effective mass of the Dirac equation. If $\sigma_H(k)=0$ passes through zero at some finite momentum, the medium is topological. The non-trivial phase $C=2$ corresponds to a gapless unidirectional photon at the boundary, dubbed the quantum gyro-electric effect (QGEE). We strongly emphasize that this model is validated by direct comparison with the supersymmetric Dirac theory for continuum fermions.}
\label{fig:Dirac_point}
\end{figure}

\subsection{Even parity bosonic phase: $C=\pm 2$}

We adopt an exactly solvable model to unravel the low energy topological physics of this phase. We let the response function be rotationally invariant $[\hat{S}_z,\mathcal{M}(\omega,k)]=0$ at all momenta, while also assuming $\varepsilon=\textrm{const.}>1$ is dielectric and the response is non-magnetic $\mu=1$. In this case, all the physics is captured by the gyrotropic coefficient $g$, which is the high-frequency analog of the DC Hall conductivity, 
\begin{equation}
g(\omega,k)=\frac{\sigma_H(k)}{\omega}, \qquad  \sigma_H(k)=\sigma_0-\sigma_2 (ka)^2.
\end{equation}
$\sigma_H(k)$ is a non-local Hall conductivity \cite{Zheng2002}. $\sigma_0$ is the static response and $\sigma_2$ characterizes the momentum dependence (scaled to the lattice constant $a$). At low energy $\omega\to 0$, this is the only admissible form of $g\neq 0$ \cite{Nikolaev1996,eroglu2010wave}. Due to the reality of the electromagnetic field $\mathcal{M}^*(-\omega,k)=\mathcal{M}(\omega,k)$, gyrotropy must always be odd in frequency $g(-\omega, k)=-g(\omega, k)$. This means a first-order pole at $\omega=0$ is permissible and corresponds to non-zero Hall conductivity $\omega g(\omega,k)=\sigma_H(k)\neq 0$. The energy density is positive definite $\bar{\mathcal{M}}=\partial_\omega (\omega\mathcal{M})=\mathrm{diag}[\varepsilon,\varepsilon,1]>0$ and non-singular at $\omega=0$.

We highlight important aspects of our model and the connections to experimentally measured gyrotropic responses. Firstly, we deal with Hermitian systems so the imaginary part of the dielectric permittivity is zero $\Im[\varepsilon]=\sigma/\omega=0$. Therefore, no dissipative currents exist in this system and the gyrotropic coefficient is related only to a dissipationless Hall current. Experimentally measured variables connecting to the gyrotropic coefficient, such as Verdet constants, are highly frequency dependent and this is consistent with our model. Furthermore, the zero frequency behavior of the gyrotropic coefficient $g=\sigma_{H}/\omega$ is in agreement with first-order poles in standard models of conductivity $\Im[\varepsilon]=\sigma/\omega$. 

Remarkably, the quadratic spatial correction to $\sigma_H$ is sufficient to describe a topological photonic phase and the continuum theory is regularized at $k\to\infty$. The interpretation is particularly simple in this context. At long wavelengths $\sigma_H(k\to 0)\to\sigma_0$, the Hall conductivity induces circulating currents of a specific handedness (clockwise or counter-clockwise), but at short wavelengths $\sigma_H(k\to \infty)\to -\sigma_2(ka)^2$, the handedness can reverse directions. We will show that when $\sigma_H$ switches sign, the phase is non-trivial.

\subsection{Bulk (bosonic Chern insulator)}

In vacuum, the photon is massless and therefore linearly dispersing $\omega=k$. This is the photonic (spin-1) equivalent of a Dirac point and arises naturally from Maxwell's equations $\mathcal{H}_0(\mathbf{k})=\mathbf{k}\cdot\mathbf{S}=k_x\hat{S}_x+k_y\hat{S}_y$, as mentioned in Sec.~\ref{sec:Continuum_Photonics}. By introducing the Hall conductivity, the linear dispersion of bulk waves fundamentally changes - a gap is formed at zero frequency $\omega=0$,
\begin{equation}
\varepsilon\omega^2(k)=k^2+\frac{\sigma^2_H(k)}{\varepsilon},
\end{equation}
where $\sigma_H$ acts identically to an effective photon mass \cite{Dunne1999,BOYANOVSKY1986483}. $\varepsilon$ governs the effective speed of light. A schematic of the vacuum and bulk dispersion is displayed in Fig.~\ref{fig:Dirac_point}.

\begin{figure}
\includegraphics[width=0.75\columnwidth]{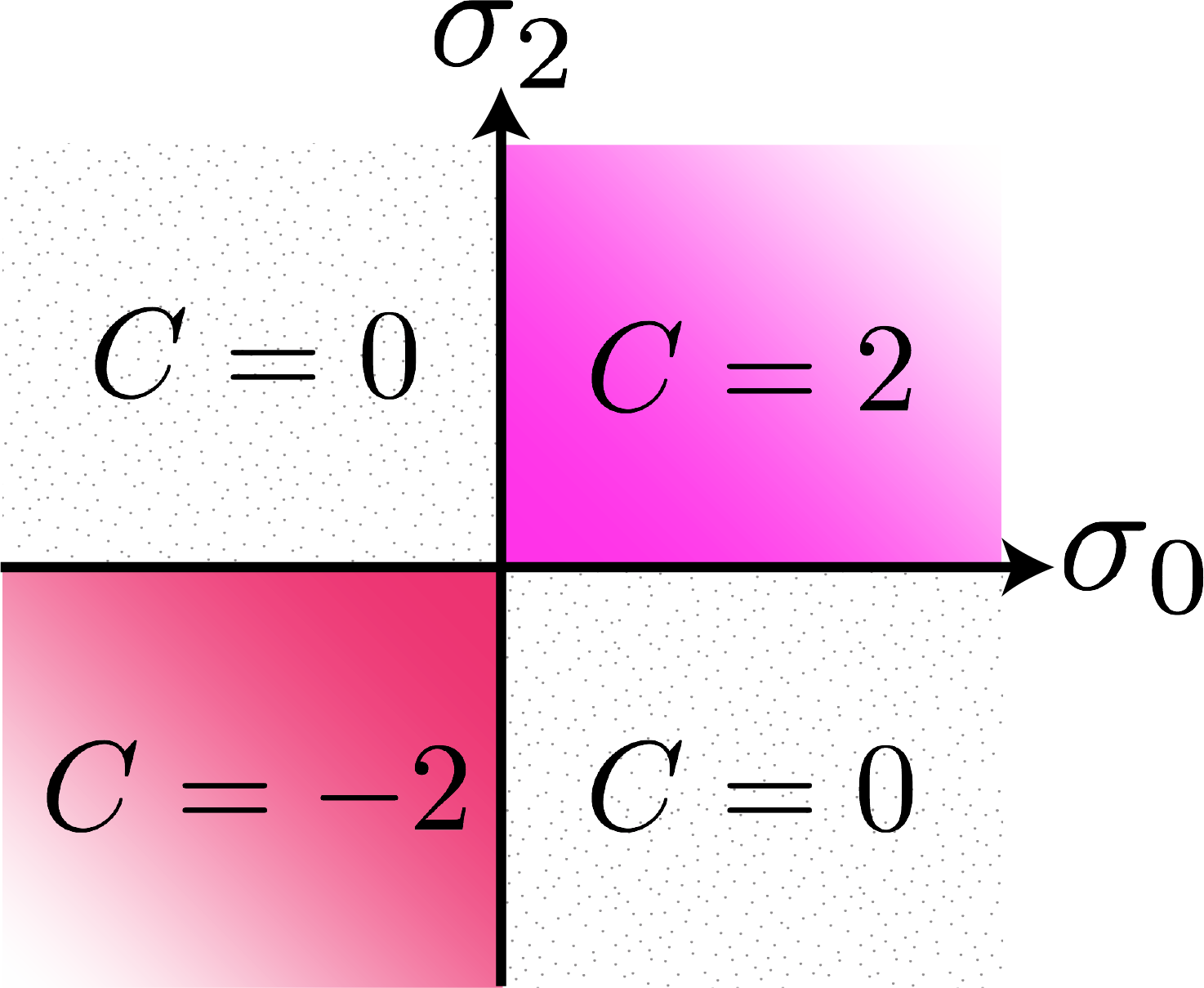}
\caption{Topological phase diagram of the non-local Hall model $\sigma_H(k)=\sigma_0-\sigma_2 (ka)^2$. $C=\mathrm{sgn}[\sigma_0]+\mathrm{sgn}[\sigma_2]$ corresponds to the Chern number of the positive frequency band $\omega>0$. The Chern number of the negative frequency band $\omega<0$ is exactly opposite $-C$. When $\sigma_0\sigma_2>0$, the photon is in a non-trivial bosonic phase $C=\pm 2$, while $\sigma_0\sigma_2<0$ is a trivial phase $C=0$. In the continuum theory, trivial and non-trivial phases can only be distinguished by incorporating non-locality $\sigma_2\neq 0$.}
\label{fig:phase_diagram}
\includegraphics[width=0.85\columnwidth]{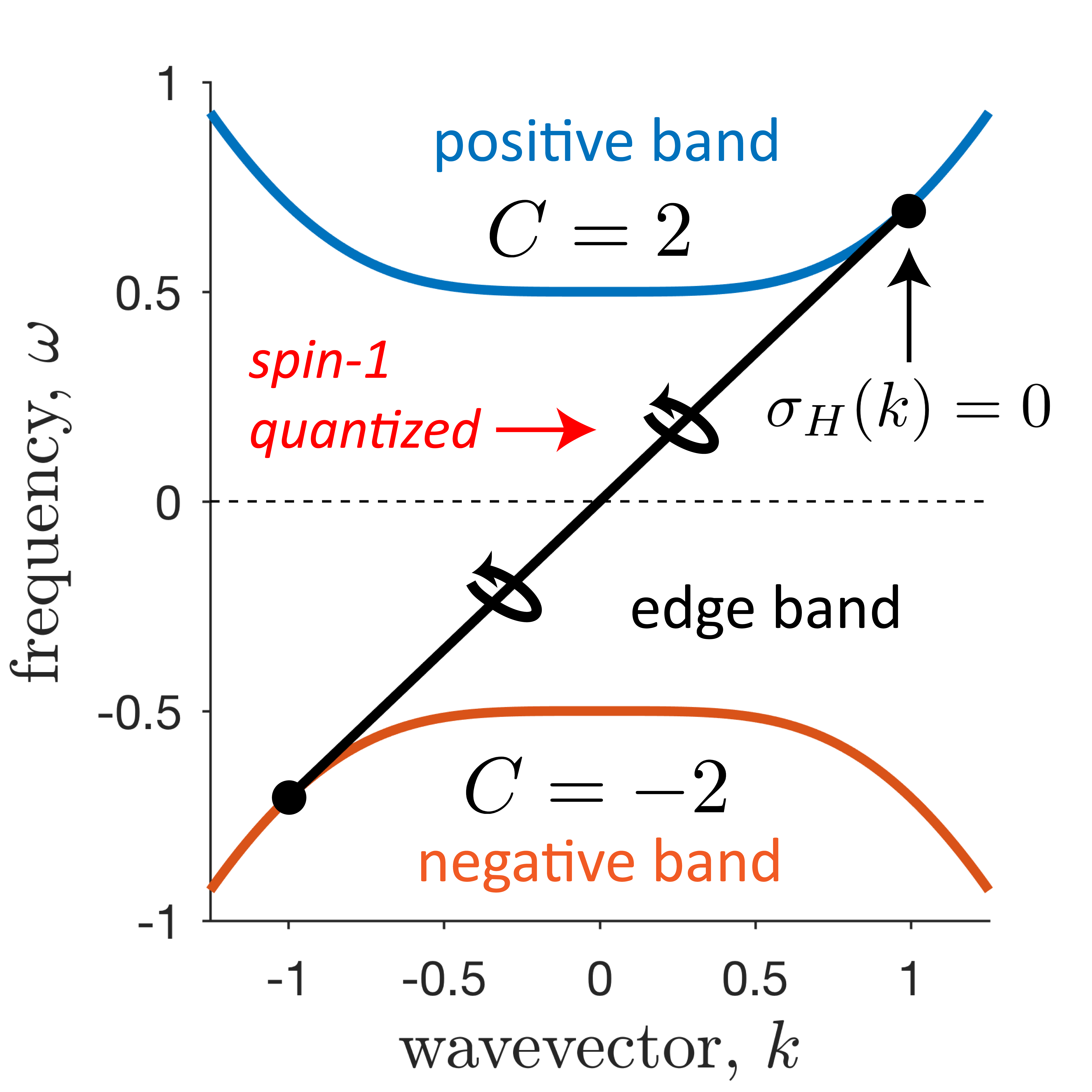}
\caption{Continuum band diagram $\omega(k)$ of the even parity $C=2$ topological bosonic phase. The negative frequency branch has a Chern number of $-2$; necessary for the total summation to vanish $2-2=0$. As an example, we have let $\sigma_0=\sigma_2a^2=1$ and $\varepsilon=2$. The unidirectional edge state is spin-1 helically quantized and touches the bulk bands precisely where the non-local Hall conductivity passes through zero $\sigma_H(k)=0$. At this point $ka=\sqrt{\sigma_0/\sigma_2}$, the edge state joins the continuum of bulk bands. Notice that no edge solution exists for $k_y\to -k_y$ and the photon is immune to backscattering.}
\label{fig:dispersion_ann}
\end{figure}

There is only one positive frequency $\omega>0$ eigenstate associated with this system and is expressed in polar coordinates as,
\begin{equation}\label{eq:Hall_eigenstate}
f_\mathbf{k}=
\begin{bmatrix}
E_x\\E_y\\H_z
\end{bmatrix}=\frac{1}{\sqrt{2\varepsilon}}\left(\frac{\sigma_H}{\varepsilon\omega}\hat{\mathbf{k}}+i\hat{\pmb{\phi}}+i\frac{k}{\omega}\hat{\mathbf{z}} \right)e^{i\phi}.
\end{equation}
$f_\mathbf{k}$ is normalized to the energy density $1=f_\mathbf{k}^\dagger\bar{\mathcal{M}}f_\mathbf{k}$ and is written in a fixed Berry gauge defined by the TAM $\hat{J}_z f_\mathbf{k}=f_\mathbf{k}$. We now show that the photon in this eigenstate exhibits transverse spin quantization at HSPs, which is independent of the chosen Berry gauge. From Eq.~(\ref{eq:Hall_eigenstate}) above, we have $\hat{S}_z f(k_p)=m(k_p) f(k_p)$ at the plasmon resonances,
\begin{equation}
m(k_p)=\frac{g(\omega(k_p),k_p)}{\varepsilon(\omega(k_p),k_p)}=\frac{\sigma_H(k_p)}{\varepsilon\omega(k_p)}=\mathrm{sgn}[\sigma_H(k_p)].
\end{equation}
Since $\varepsilon$ is a constant, the eigenvalues are determined solely by the long and short wavelength behavior of the Hall conductivity, $m(0)=\mathrm{sgn}[\sigma_0]$ and $m(\infty)=-\mathrm{sgn}[\sigma_2]$, giving a Chern number of,
\begin{equation}
C=m(0)-m(\infty)=\mathrm{sgn}[\sigma_0]+\mathrm{sgn}[\sigma_2].
\end{equation}
A non-trivial phase corresponds to $\sigma_0\sigma_2>0$, which can be $C=\pm 2$ depending on the signs of $\sigma_0$ and $\sigma_2$. This is the simplest realization of a bosonic Chern insulator \cite{Jotzu2014}. It is equally important to note that $\sigma_0\sigma_2<0$ corresponds to a trivial phase $C=0$. Distinguishing between trivial $C=0$ and non-trivial $C=\pm 2$ phases is only possible by incorporating non-locality $\sigma_2\neq 0$. A topological phase diagram of this system is presented in Fig.~\ref{fig:phase_diagram}.

In the non-trivial phase, there is a point where the Hall conductivity $\sigma_H(k)=0$ passes through zero - precisely at $ka=\sqrt{\sigma_0/\sigma_2}$. The zero must occur for the spin to change handedness and can only be removed by a topological phase transition. This also puts an approximate bound on the Hall parameters. As long as $\sqrt{\sigma_0/\sigma_2}\ll\pi$ the continuum theory is valid and the zero occurs within the Brillouin zone. As an aside, we note the negative frequency band $\omega<0$ has a Chern number of $-C$ - exactly opposite of the positive band. This is necessary to ensure the summation over all bands vanishes $\sum_n C_n=0$.

\subsection{Edge (quantum gyro-electric effect)}

\begin{figure*}
\includegraphics[width=0.9\linewidth]{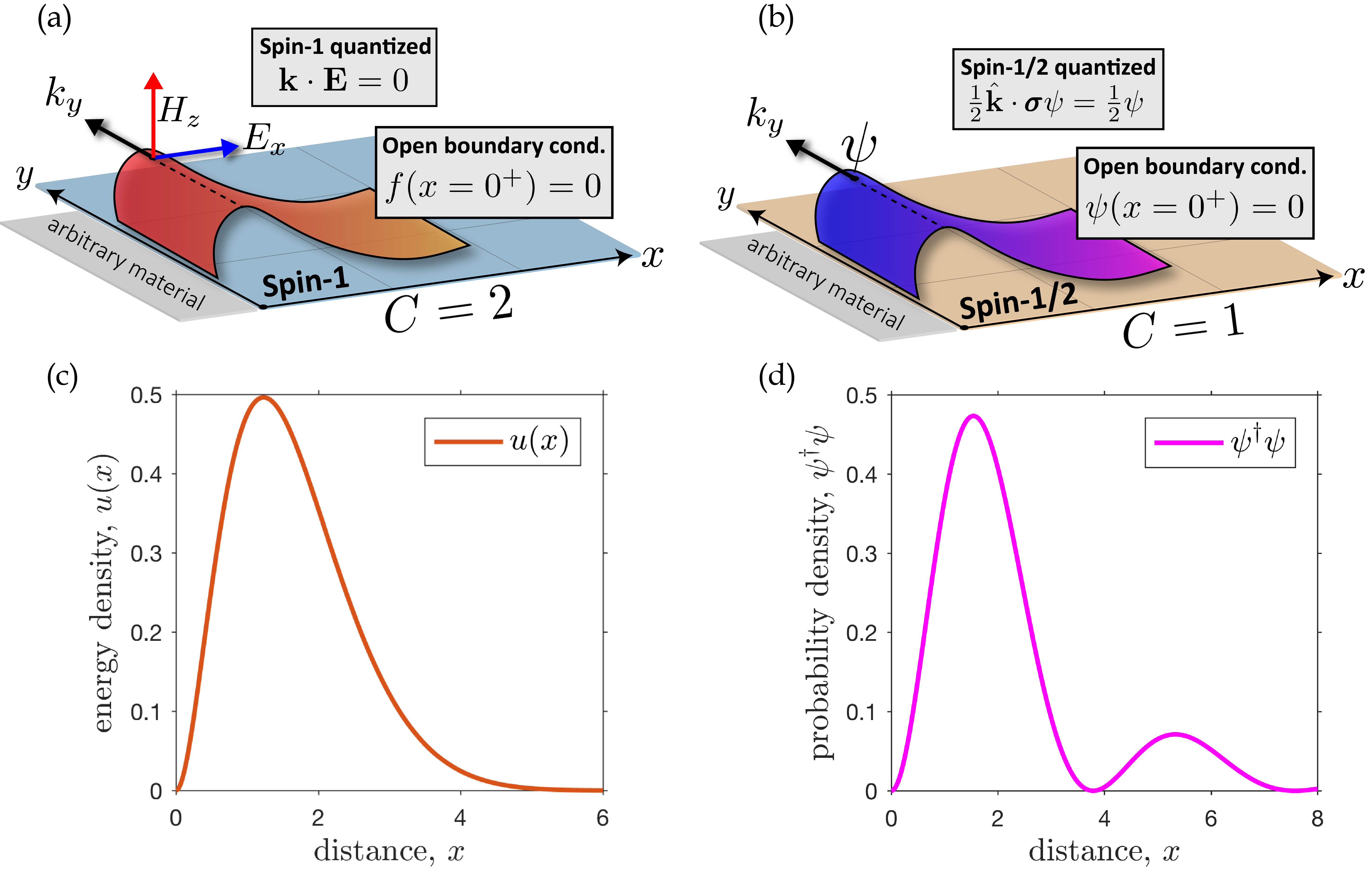}
\caption{\textbf{(a)} Topological edge state of the even parity $C=2$ bosonic phase. The photon is spin-1 helically quantized $\mathbf{k}\cdot\mathbf{E}=0$ and satisfies open boundary conditions at the interface $f(x=0^+)=0$. This ensures the edge state is immune to boundary defects and can exist at any interface - even vacuum. \textbf{(b)} Topological edge state of the $C=1$ fermionic phase. Like the photon, the electron is spin-\textonehalf{} helically quantized $\frac{1}{2}\hat{\mathbf{k}}\cdot\pmb{\sigma}\psi=\frac{1}{2}\psi$ and satisfies open boundary conditions $\psi(x=0^+)=0$. \textbf{(c)} Normalized energy density $u(x)=f^\dagger\bar{\mathcal{M}}f$ of the unidirectional photon as a function of distance $x$, at a momentum of $k_y=0.5$. As an example we have let $\sigma_0=\sigma_2a^2=1$ and $\varepsilon=2$. Notice the fields are identically zero at $x=0$ and the edge state exists at the boundary of any interface. \textbf{(d)} Probability density $\psi^\dagger\psi$ of the electronic edge state, where we have let $\Lambda_0=\Lambda_2a^2=1$ and $v=0.5$ as an example. The probability density is evaluated for a momentum of $k_y=0.5$.}
\label{fig:boundary_conditions}
\end{figure*}

Finally, we analyze the unique edge state of this bosonic phase, which has no counterpart in traditional surface photonics - such as plasmon polaritons, Tamm states, Dyakonov or Zenneck waves \cite{Takayama2008}. This is because topological boundary conditions are captured by non-local (spatially and temporally dispersive) optical constants. In conventional problems, non-locality introduces additional boundary conditions (ABCs) \cite{agranovich2013crystal,Halevi1984} which need to be satisfied to uniquely determine the electromagnetic field. Our newly discovered photonic edge state is fundamentally different in this context. The behavior of the field outside the medium $x<0$ becomes irrelevant due to topological \textbf{open boundary conditions} $f(x=0^+)=0$ \cite{Shenoy2012,Shen2010,SHEN2011,gelfand2000calculus}. Open boundary conditions are commonly encountered in topological electronics \cite{Hatsugai1993,avila_topological_2013} but is surprising when dealing with photons. To be localized at the edge, all components of the field must decay into the bulk $f(x=\infty)=0$ as $x\to\infty$ and simultaneously disappear on the edge. The exact bulk and edge dispersion is plotted in Fig.~\ref{fig:dispersion_ann} and a diagram of the topological edge state is displayed in Fig.~\ref{fig:boundary_conditions}(a). We strongly emphasize that these special solutions point to the first unified topological theory of Maxwell-bosons and Dirac-fermions. 

The specific phase $C=\pm 2$ will determine if the unidirectional photon is forward or backward propagating; forward for $C=2$ and backward for $C=-2$. We stress again that for either $C=\pm 2$, there is only one bosonic solution at the boundary - not two. In either case, the solution in the $x>0$ half-space has a similar form $f_\pm(x,y)=f_\pm(x)e^{ik_y y}$. Inserting into the wave equation and applying open boundary conditions, the topological edge state emerges,
\begin{subequations}
\begin{equation}
\omega_\pm=\pm\frac{k_y}{\sqrt{\varepsilon}}, \qquad -\sqrt{\frac{\sigma_0}{\sigma_2}}< k_ya<\sqrt{\frac{\sigma_0}{\sigma_2}},
\end{equation}
\begin{equation}
f_\pm(x)=\begin{bmatrix}
E_x\\E_y\\H_z
\end{bmatrix}_\pm=f_0(\hat{\mathbf{x}}\mp\sqrt{\varepsilon}\hat{\mathbf{z}})\left(e^{-\eta_1 x}-e^{-\eta_2 x}\right).
\end{equation}
\end{subequations}
A solution only exists in the non-trivial phase $\sigma_0\sigma_2>0$, confirming our theory. Notice the group velocity $v_\pm =\partial \omega_\pm/\partial k_y=\pm 1/\sqrt{\varepsilon}$ is constant and the edge state can propagate in opposite directions depending on the phase $C=\pm 2$. Moreover, since no solution exists for $k_y\to -k_y$, the photon is immune to backscattering. The decay lengths $\eta_1$ and $\eta_2$ are found from the two quadratic roots,
\begin{equation}
\eta_{1,2} =\frac{1}{2a^2|\sigma_2|}\left\{\sqrt{\varepsilon}\pm\sqrt{\varepsilon+4\sigma_2a^2[\sigma_2(k_ya)^2-\sigma_0]}\right\},
\end{equation}
which determine the degree of confinement at a particular wavevector $k_y$. A plot of the electromagnetic energy density is presented in Fig.~\ref{fig:boundary_conditions}(c). Intriguingly, the field is completely transverse polarized $\hat{\mathbf{k}}\cdot\mathbf{E}=E_y=0$ and helically quantized along the direction of momentum $k_y$,
\begin{equation}
\frac{f_\pm^\dagger\hat{S}_yf_\pm}{f_\pm^\dagger\bar{\mathcal{M}}f_\pm} =v_\pm=\pm\frac{1}{\sqrt{\varepsilon}}.
\end{equation}
$\hat{\mathbf{k}}\cdot\mathbf{S}=\hat{S}_y$ is the spin-1 helicity operator and quantization lies in the $x$-$y$ plane. Consequently, the $C=2$ phase corresponds to a massless (linearly dispersing) ``spin-up'' photon while the $C=-2$ phase is a counter-propagating ``spin-down'' photon. Note, the edge wave is $\hat{S}_y$ helically quantized for \textbf{all momenta} and is distinct from transverse $\hat{S}_z$ quantization of the bulk waves, which only occurs at HSPs.

\subsection{Anomalous displacement currents}

We also discover an anomalous edge current \cite{Nagaosa2010} propagating parallel to the interface,
\begin{equation}
J_y(x,y)=\mp \sqrt{\varepsilon} f_0(\eta_1 e^{-\eta_1 x}-\eta_2 e^{-\eta_2 x})e^{ik_y y}.
\end{equation}
The displacement current is induced by the non-local Hall conductivity,
\begin{equation}
J_y=-\partial_x H_z=-\left(\sigma_0+\sigma_2a^2\nabla^2\right)E_x,
\end{equation}
and is highly conductive near the interface $J_y(x=0^+)\neq 0$. However, a compensating current is generated in the bulk $x>0$, such that the total induced charge is identically zero $\int_{0^+}^\infty J_y(x) dx =H_z(0^+)-H_z(\infty)=0$. Notice that charge neutrality is only guaranteed by the open boundary condition $H_z(x=0^+)=0$, providing a profound physical basis for topological protection. The photonic edge state must exist if the medium is to remain neutral - there is no other option.

\begin{figure}
\includegraphics[width=0.9\columnwidth]{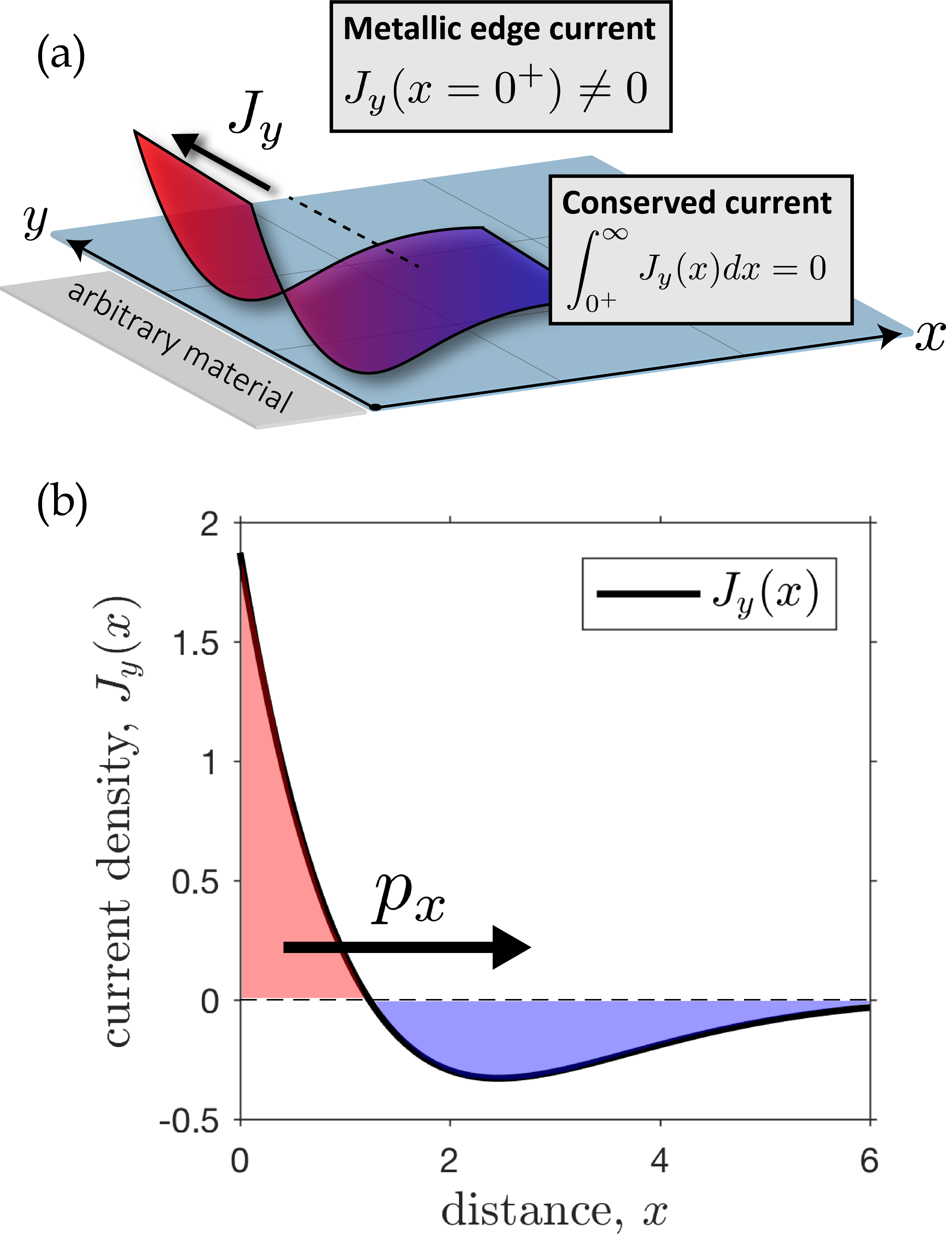}
\caption{\textbf{(a)} Anomalous displacement current at the edge of the topological photonic medium. \textbf{(b)} Real current density $J_y(x)$ as a function of distance $x$, for a momentum of $k_y=0.5$. We have let $\sigma_0=\sigma_2a^2=1$ and $\varepsilon=2$ as an example. The displacement current is generated by the non-local Hall conductivity and is highly metallic near the interface $J_y(x=0^+)\neq 0$. However, the total current is conserved $\int_{0^+}^\infty J_y(x) dx =0$ which is clear from the positive (red) and negative (blue) charge density. Since the net charge is zero, this phenomenon can be interpreted as a propagating dipole bound to the edge of the material - with an intrinsic dipole moment $p_x=\int_{0^+}^\infty x\rho(x) dx$.}
\label{fig:current_diagram}
\end{figure}

From the continuity equation $\omega \rho=k_y J_y$, this phenomenon can also be interpreted as a propagating dipole bound to the edge of the material,
\begin{equation}
p_x=\int_{0^+}^\infty x\rho(x) dx =-\varepsilon\int_{0^+}^\infty E_x(x) dx=\varepsilon f_0(\eta_1^{-1}-\eta_2^{-1}),
\end{equation}
with an intrinsic dipole moment $p_x$ normal to the interface. The intriguing connection to the parity anomaly will be discussed in a future paper \cite{Haldane1988,Fradkin1986}. The dipole is continuous $x\rho(x)$ and shields the electromagnetic field between regions of positive and negative charge density. This unusual effect allows highly confined photons to propagate at the boundary unimpeded, impervious to defects. A visualization of the anomalous current is displayed in Fig.~\ref{fig:current_diagram}.

\section{Dirac-Maxwell supersymmetry: Direct correspondence between fermionic and bosonic phases}\label{sec:Dirac_Maxwell}

To validate our predictions of the Maxwell theory, we solve the equivalent continuum Dirac theory. These results present a unified topological field theory of Maxwell-bosons and Dirac-fermions. It also highlights the one-to-one correspondence between even parity bosonic phases $C=\pm 2$ and traditional fermionic phases $C=\pm 1$. Interestingly, the equivalent 2D Dirac theory is a supersymmetric partner of the 2D Maxwell theory \cite{Dunne1999,BOYANOVSKY1986483,Ziegler1998,gates2001}. The continuum Dirac Hamiltonian is given succinctly as,
\begin{equation}
H(\mathbf{k})=v(k_x\sigma_x+k_y\sigma_y)+\Lambda(k)\sigma_z,
\end{equation}
where $\sigma_i$ are the SU(2) Pauli matrices. The dispersion relation of the positive energy state $E>0$ is found as,
\begin{equation}
E^2(k)=v^2 k^2+\Lambda^2(k),
\end{equation}
and $\Lambda(k)=\Lambda_0-\Lambda_2 (ka)^2$ is a spatially dispersive Dirac mass \cite{Shenoy2012,Shen2010,SHEN2011}. $v$ being the Fermi velocity. Again, we include quadratic $k$ dependence for proper regularization at $k\to \infty$. This direct correspondence makes our earlier claim, the necessity of non-locality (momentum dependence), very clear. 

The Hamiltonian possesses rotational symmetry, which is generated by the spin-\textonehalf{} operator $\hat{S}_z=\frac{1}{2}\sigma_z$. This is evidently a fermionic representation $\mathcal{R}(2\pi)=\exp[i2\pi \hat{S}_z]=-\mathds{1}_2$. Furthermore, we can prove transverse spin-\textonehalf{} quantization at HSPs, $m(0)=\frac{1}{2}\mathrm{sgn}[\Lambda_0]$ and $m(\infty)=-\frac{1}{2}\mathrm{sgn}[\Lambda_2]$. This should be contrasted with our results for integer spin quantization of the 2D bosonic phase in Fig.~(\ref{fig:quantization}). We obtain a Chern invariant of,
\begin{equation}
C=m(0)-m(\infty)=\frac{1}{2}\left(\mathrm{sgn}[\Lambda_0]+\mathrm{sgn}[\Lambda_2]\right).
\end{equation}
The phase is only non-trivial $C=\pm 1$ when $\Lambda_0\Lambda_2>0$, necessitating a zero in the effective mass $\Lambda(k)=0$ - precisely at $ka=\sqrt{\Lambda_0/\Lambda_2}$. This is the simplest realization of a fermionic Chern insulator \cite{bernevig2013topological}. Notice that in two dimensions, the Hall conductivity $\sigma_H$ for the photon plays an analogous role as the Dirac mass $\Lambda$ for the electron.

The electronic edge state has a similar interpretation as the photon, but for spin-\textonehalf{} particles. Imposing open boundary conditions, a unidirectional edge state is revealed $\psi_\pm(x,y)=\psi_\pm(x)e^{ik_y y}$,
\begin{equation}
\psi_\pm(x)=\psi_0\begin{bmatrix}
1\\ \pm i
\end{bmatrix}(e^{-\eta_1 x}-e^{-\eta_2 x}),
\end{equation}
corresponding to a $C=\pm 1$ phase. The decay lengths have an identical form,
\begin{equation}
\eta_{1,2} =\frac{1}{2a^2|\Lambda_2|}\left\{v\pm\sqrt{v^2+4\Lambda_2a^2[\Lambda_2(k_ya)^2-\Lambda_0]}\right\},
\end{equation}
and the edge state is massless $E_\pm=\pm v k_y$, propagating in opposite directions depending on the phase. Furthermore, the edge state is spin-\textonehalf{} helically quantized $\frac{1}{2}\hat{\mathbf{k}}\cdot\pmb{\sigma}\psi_\pm= \frac{1}{2}\sigma_y \psi_\pm =\pm\frac{1}{2}\psi_\pm$, equating to a spin-up or spin-down electron for $C=1$ or $C=-1$ respectively. The striking similarity of Maxwell-bosons and Dirac-fermions is shown in Fig.~\ref{fig:boundary_conditions}. A diagram of the electronic topological edge state is displayed in Fig.~\ref{fig:boundary_conditions}(b) and a plot of the probability density is given in Fig.~\ref{fig:boundary_conditions}(d).

\section{Experimental Search and Conclusions}\label{sec:Conclusions}

The standard approach of characterizing TRS broken fermionic phases is DC transport measurements (charge and spin Hall conductivity) or Faraday rotation angles at THz frequencies. It should be noted that the former measurement gives information at zero frequency and zero momentum $\sigma_H(0,0)$ whereas the latter experiment characterizes matter at finite frequency but close to zero momentum $\sigma_H(\omega,0)$. Our predictions of photon spin quantization and bosonic phases are fundamentally tied to $\sigma_H(\omega,k)$. This is a formidable challenge and therefore, for completeness, we suggest two new experimental approaches.

\subsection{Momentum-resolved electron energy loss spectroscopy of gyrotropic plasmas}

The high-frequency $\omega >0$ and sub-wavelength $k>0$ properties of matter can be probed by momentum-resolved electron energy loss spectroscopy (k-EELS) \cite{Shekhar2017}. Here, highly energetic electrons pass through matter and their energy loss, as well as their momentum loss, is measured to understand the bulk light-matter excitations. Fundamentally different from conventional STEM-EELS \cite{scholl_quantum_2012}, this approach can also give insight into high momentum waves through scattering angle measurement of electrons passing through matter. We anticipate non-local gyrotropic plasmas to be ideal candidates for topological bosonic phases of matter and probing with k-EELS.

\subsection{Cold atom near-field probes of non-local optical conductivity}

Dynamical (high-frequency) conductivity is regularly studied by conventional tools such as ellipsometry and Faraday rotation using incident optical beams. However, the momentum carried by light waves is negligible compared to the Fermi momentum of electrons. Therefore, the large momentum $k>0$ behavior of the conductivity requires fundamentally new probes. One approach is to use spontaneous emission from cold atoms in the near-field to investigate deep sub-wavelength response parameters of our predicted bosonic phases of matter. This is feasible since the GHz splitting in Rydberg atoms \cite{Saffman2010} and low-frequency gyrotropic response in systems such as quantum wells are comparable \cite{Jin2016}. Recent work has shown trapping of cold atoms near photonic nanostructures \cite{Kimble2015} - a promising route for probing topological properties of matter.

\subsection{Conclusions}

In summary, we have developed the complete continuum field theory describing all 2+1D topological bosonic phases of the photon; incorporating both temporal and spatial dispersion as a necessary generalization. The topological phases are intimately connected to photon spin-1 quantization, with non-locality being imperative to properly characterize the high-$k$ global behavior. Two unique bosonic phases are predicted - an even parity phase $C=\pm 2$ which is understood in interacting bosonic systems, and an odd parity phase $C=\pm 1$ which has no immediate interpretation but presents possibly unexplored physics. We have studied the even parity phase $C=\pm 2$ utilizing a non-local Hall conductivity and reveal a single topologically protected unidirectional photon at the edge. This photon is helically quantized (spin-1), immune to backscattering, defects, and exists at the boundary of the $C=\pm 2$ bosonic phase and any interface - even vacuum. To validate our theory, we have compared all the low energy Maxwell phenomena to its supersymmetric Dirac counterpart, confirming that even parity bosonic phases $C=\pm 2$ are the exact analogue of traditional fermionic phases $C=\pm 1$.

\section*{Acknowledgements}

This research was supported by the Defense Advanced Research Projects Agency (grant \#: N66001-17-1-4048) and the National Science Foundation (EFMA-1641101). We also thank Dr. Cristian Cortes for insightful discussions.

\appendix

\section*{Appendix}

\section{2+1D electromagnetic Lagrangian}\label{app:2DElectro}

To understand the two dimensional behavior of photons \cite{Dunne1999}, we start with the electromagnetic Lagrangian coupled to a conserved current $\partial_\mu J^\mu=0$,
\begin{equation}
\mathcal{L}=-\frac{1}{4}F^{\mu\nu}F_{\mu\nu}-A_\mu J^\mu, \qquad F^{\mu\nu}=\partial^{\mu}A^\nu-\partial^{\nu}A^\mu,
\end{equation}
which is exact in any space-time dimension. The conservation of charge ensures the action $S=\int d^dxdt \mathcal{L}$ is gauge invariant, where $d$ is the spatial dimension. For $d=2$, the motion of charge is restricted to the $x$-$y$ plane,
\begin{equation}
J^\mu=(\rho,J_x,J_y), \qquad \dot{\rho}+\partial_i J^i=0.
\end{equation}
Similarly, planar currents restrict the spatial degrees of freedom of the gauge potential $A^\mu$,
\begin{equation}
A^\mu=(\phi,A_x,A_y).
\end{equation}
This implies there are only 2 components of the electric field and 1 for the magnetic field,
\begin{equation}
E_i=-\partial_i\phi-\dot{A}_i, \qquad B_z=\epsilon^{ij}\partial_iA_j=\partial_xA_y-\partial_y A_x,
\end{equation}
such that exclusively transverse-magnetic (TM) waves propagate within a 2D material. This makes physical sense since the circulation of currents can only generate magnetic fields in a single $\hat{\mathbf{z}}$ direction. Note that $\epsilon_{ij}=-\epsilon_{ji}$ is the 2D antisymmetric matrix and should not be confused with the permittivity.

Varying the action with respect to $A^\mu$, we arrive at the familiar equations of motion,
\begin{equation}\label{eq:current}
\partial_\mu F^{\mu\nu}=J^{\nu}, \qquad \tilde{F}^{\mu}=\frac{1}{2}\epsilon^{\mu\nu\rho}F_{\nu\rho}, \qquad \partial_\mu\tilde{F}^{\mu}=0.
\end{equation}
Notice the dual equation $\tilde{F}_\mu$ is slightly different in 2D, which arises from the fact there are only 3 unique components of the electromagnetic field. We can express the equations of motion directly in terms of $E_i$ and $B_z$,
\begin{equation}
\partial_i E^i=\rho, \qquad \epsilon_{ij}\partial^j B_z-\dot{E}_i=J_i, \qquad \dot{B}_z+\epsilon^{ij}\partial_iE_j=0.
\end{equation}
These are precisely Maxwell's equations in 2D.

We are most interested with the response of a bulk 2D material so it is convenient to represent the induced charges in terms of the polarization $P_i$ and magnetization $M_z$ densities,
\begin{equation}
\rho=-\partial_i P^i, \qquad J_i=\dot{P}_i+\epsilon_{ij}\partial^j M_z.
\end{equation}
Substituting into the equations of motion, we define the electric $D_i$ and magnetic $B_z$ displacement fields as,
\begin{equation}
\dot{D}_i-\epsilon_{ij}\partial^jH_z=0, \qquad \dot{B}_z+\epsilon^{ij}\partial_iE_j=0,
\end{equation}
which is simply the linear superposition of,
\begin{equation}
D_i=E_i+P_i, \qquad  B_z=H_z+M_z.
\end{equation}
The wave equation in Eq.~(\ref{eq:wave_equation}) follows immediately after substituting for the linear response function $\mathcal{M}$ and defining the column vector $f=[ E_x ~ E_y ~ H_z]^\intercal$ for the TM state.

\section{Electromagnetic Hamiltonian and polariton eigenstates}\label{app:Oscillators}

Here, we show that the response function $\mathcal{M}$ is derived form a first-order in time Hamiltonian. Utilizing the decomposition in Eq.~(\ref{eq:Response}), we expand in terms of 3-component oscillator variables $\psi_\alpha$ by defining,
\begin{equation}\label{eq:oscillator_coupling}
\psi_\alpha=\frac{\mathcal{C}_{\alpha\mathbf{k}}f}{\omega-\omega_{\alpha\mathbf{k}}}, \qquad \omega \psi_\alpha=\omega_{\alpha\mathbf{k}}\psi_\alpha+\mathcal{C}_{\alpha\mathbf{k}}f,
\end{equation}
which is first-order in time. Similarly, we back-substitute Eq.~(\ref{eq:Response}) into Eq.~(\ref{eq:wave_equation}) to obtain,
\begin{equation}\label{eq:field_coupling}
\omega f=\left[\mathcal{H}_0(\mathbf{k})+\sum_\alpha \omega_{\alpha\mathbf{k}} ^{-1}\mathcal{C}^\dagger_{\alpha\mathbf{k}} \mathcal{C}_{\alpha\mathbf{k}}\right]f+\sum_\alpha \mathcal{C}^\dagger_{\alpha\mathbf{k}} \psi_\alpha.
\end{equation}
The first term represents the vacuum equations and self-energy of the electromagnetic field, while the second is the linear coupling to the oscillators. By combining Eq.~(\ref{eq:oscillator_coupling}) and (\ref{eq:field_coupling}) into a single algebraic matrix, the complete electromagnetic Hamiltonian emerges,
\begin{equation}
H(\mathbf{k})=\begin{bmatrix}
\mathcal{H}_0(\mathbf{k})+\sum_\alpha \omega_{\alpha\mathbf{k}} ^{-1}\mathcal{C}^\dagger_{\alpha\mathbf{k}} \mathcal{C}_{\alpha\mathbf{k}} & ~~\mathcal{C}^\dagger_{1\mathbf{k}}~~ & ~~\mathcal{C}^\dagger_{2\mathbf{k}}~~ & \ldots \\
\mathcal{C}_{1\mathbf{k}} & \omega_{1\mathbf{k}} & 0 & \ldots\\
\mathcal{C}_{2\mathbf{k}} & 0 & \omega_{2\mathbf{k}} & \ldots\\
\vdots & \vdots & \vdots & \ddots 
\end{bmatrix}.
\end{equation}
The Hermitian equation $Hu=\omega u $ characterizes the dynamics of the entire electromagnetic problem in a 2D material. $u$ constitutes the cumulative state vector of the photon + all oscillator degrees of freedom,
\begin{equation}
u=\begin{bmatrix}
f & \psi_1 & \psi_2 & \ldots
\end{bmatrix}^\intercal.
\end{equation}
Notice that contraction of $u$ naturally reproduces the energy density upon summation over all degrees of freedom,
\begin{equation}
u^\dagger u=f^\dagger f+f^\dagger \sum_\alpha\frac{\mathcal{C}^\dagger_{\alpha\mathbf{k}} \mathcal{C}_{\alpha\mathbf{k}}}{(\omega-\omega_{\alpha\mathbf{k}})^2} f= f^\dagger \bar{\mathcal{M}}f,
\end{equation}
with $\bar{\mathcal{M}}=\partial_\omega(\omega\mathcal{M})>0$ always positive definite.

Eigenstates of the Hamiltonian are collective excitations of oscillators coupled to the electromagnetic field, 
\begin{equation}
H_\mathbf{k}u_{n\mathbf{k}}=\omega_{n\mathbf{k}}u_{n\mathbf{k}},
\end{equation}
and are manifestly bosonic quasiparticles. These are the $n$ non-trivial roots of the characteristic equation, 
\begin{equation}
\det[\mathcal{H}_0(\mathbf{k})-\omega\mathcal{M}(\omega,\mathbf{k})]=0, \qquad \omega=\omega_n(\mathbf{k}),
\end{equation}
which generates the eigenenergies at any particular momenta. Normalization of each mode is given concisely as $1=f^\dagger_{n\mathbf{k}}\bar{\mathcal{M}}(\omega_{n\mathbf{k}},\mathbf{k})f_{n\mathbf{k}}$.

\section{Continuum regularization}\label{app:Continuum}

To adequately describe a continuum topological field theory, the Hamiltonian must approach a directionally independent value in the asymptotic limit $\lim_{k\to\infty}H(\mathbf{k})\to H(k)$, such that the system is connected at infinity \cite{Ryu2010}. This is the continuum equivalent of a periodic boundary condition since all limits at $k\to \infty$ are mapped into a single point (i.e. one-point compactification). We can prove the Chern number is quantized by analyzing the Berry phase over all momentum. Continuum regularization necessitates the following condition,
\begin{equation}
\oint _{\infty}\mathbf{A}_n\cdot d\mathbf{k}=-2\pi\sum_i  p_i +\iint_{\mathbb{R}^2}F_n ~d^2\mathbf{k} =2\pi p,
\end{equation}
with $p$ and $p_i\in \mathbb{Z}$ an integer. Here, $\mathbf{A}_n(\mathbf{k})=-iu_{n\mathbf{k}}^\dagger\partial_\mathbf{k} u_{n\mathbf{k}}$ is the Berry connection of any particular eigenstate and $F_n(\mathbf{k})=\hat{\mathbf{z}}\cdot[\partial_\mathbf{k}\times\mathbf{A}_n(\mathbf{k})]$ is the Berry curvature. The path integral is performed over a closed loop at infinity $k=\infty$, which is equivalent to the Berry flux over all momentum space $\mathbb{R}^2$ minus any singular points in the connection. $p_i$ label these singular points of the Berry connection $\mathbf{A}_n(\mathbf{k}_i)$ which contribute an integer unit of Berry flux at a particular momentum $\mathbf{k}_i$. The Chern number $C_n\in \mathbb{Z}$ is the summation over all such singularities,
\begin{equation}\label{eq:berry_constraint}
C_n= p+\sum_i p_i=\frac{1}{2\pi}\iint_{\mathbb{R}^2}F_n ~d^2\mathbf{k}.
\end{equation}
For Eq.~(\ref{eq:berry_constraint}) to hold, we see that the eigenstates must approach a directionally independent value in the asymptotic limit, up to a possible U(1) gauge,
\begin{equation}
\lim_{k\to\infty}u_n(\mathbf{k}) \to u_n(k)\exp[i\chi_n(\mathbf{k})].
\end{equation}
When this is the case, the closed loop at infinity $k=\infty$ is determined purely by the gauge,
\begin{equation}
\oint _{\infty}\mathbf{A}_n\cdot d\mathbf{k}=\oint _{\infty}\partial_\mathbf{k}\chi_n\cdot d\mathbf{k}=\chi_n|_0^{2\pi}=2\pi p,
\end{equation}
which is guaranteed to be an integer multiple of $2\pi$. Hence, Chern numbers are quantized.

For completeness, we note that the Berry connection can be simplified slightly to,
\begin{equation}
\mathbf{A}_n(\mathbf{k})=-if^\dagger_{n\mathbf{k}}\bar{\mathcal{M}}(\omega_{n\mathbf{k}},\mathbf{k})\partial_\mathbf{k} f_{n\mathbf{k}}+f^\dagger_{n\mathbf{k}}\pmb{\mathcal{A}}(\omega_{n\mathbf{k}},\mathbf{k}) f_{n\mathbf{k}},
\end{equation}
where $\pmb{\mathcal{A}}$ is the Berry connection arising from the oscillators,
\begin{equation}
\pmb{\mathcal{A}}(\omega,\mathbf{k})=-i\sum_\alpha \frac{\mathcal{C}^\dagger_{\alpha\mathbf{k}}\partial_\mathbf{k}\mathcal{C}_{\alpha\mathbf{k}}}{(\omega-\omega_{\alpha\mathbf{k}})^2}.
\end{equation}

\bibliography{spin_quant.bib}

\begin{thebibliography}{77}%
\makeatletter
\providecommand \@ifxundefined [1]{%
 \@ifx{#1\undefined}
}%
\providecommand \@ifnum [1]{%
 \ifnum #1\expandafter \@firstoftwo
 \else \expandafter \@secondoftwo
 \fi
}%
\providecommand \@ifx [1]{%
 \ifx #1\expandafter \@firstoftwo
 \else \expandafter \@secondoftwo
 \fi
}%
\providecommand \natexlab [1]{#1}%
\providecommand \enquote  [1]{``#1''}%
\providecommand \bibnamefont  [1]{#1}%
\providecommand \bibfnamefont [1]{#1}%
\providecommand \citenamefont [1]{#1}%
\providecommand \href@noop [0]{\@secondoftwo}%
\providecommand \href [0]{\begingroup \@sanitize@url \@href}%
\providecommand \@href[1]{\@@startlink{#1}\@@href}%
\providecommand \@@href[1]{\endgroup#1\@@endlink}%
\providecommand \@sanitize@url [0]{\catcode `\\12\catcode `\$12\catcode
  `\&12\catcode `\#12\catcode `\^12\catcode `\_12\catcode `\%12\relax}%
\providecommand \@@startlink[1]{}%
\providecommand \@@endlink[0]{}%
\providecommand \url  [0]{\begingroup\@sanitize@url \@url }%
\providecommand \@url [1]{\endgroup\@href {#1}{\urlprefix }}%
\providecommand \urlprefix  [0]{URL }%
\providecommand \Eprint [0]{\href }%
\providecommand \doibase [0]{http://dx.doi.org/}%
\providecommand \selectlanguage [0]{\@gobble}%
\providecommand \bibinfo  [0]{\@secondoftwo}%
\providecommand \bibfield  [0]{\@secondoftwo}%
\providecommand \translation [1]{[#1]}%
\providecommand \BibitemOpen [0]{}%
\providecommand \bibitemStop [0]{}%
\providecommand \bibitemNoStop [0]{.\EOS\space}%
\providecommand \EOS [0]{\spacefactor3000\relax}%
\providecommand \BibitemShut  [1]{\csname bibitem#1\endcsname}%
\let\auto@bib@innerbib\@empty
\bibitem [{\citenamefont {Klitzing}\ \emph {et~al.}(1980)\citenamefont
  {Klitzing}, \citenamefont {Dorda},\ and\ \citenamefont
  {Pepper}}]{Klitzing1980}%
  \BibitemOpen
  \bibfield  {author} {\bibinfo {author} {\bibfnamefont {K.~v.}\ \bibnamefont
  {Klitzing}}, \bibinfo {author} {\bibfnamefont {G.}~\bibnamefont {Dorda}}, \
  and\ \bibinfo {author} {\bibfnamefont {M.}~\bibnamefont {Pepper}},\
  }\bibfield  {title} {\enquote {\bibinfo {title} {New method for high-accuracy
  determination of the fine-structure constant based on quantized hall
  resistance},}\ }\href {\doibase 10.1103/PhysRevLett.45.494} {\bibfield
  {journal} {\bibinfo  {journal} {Phys. Rev. Lett.}\ }\textbf {\bibinfo
  {volume} {45}},\ \bibinfo {pages} {494--497} (\bibinfo {year}
  {1980})}\BibitemShut {NoStop}%
\bibitem [{\citenamefont {Laughlin}(1981)}]{Laughlin1981}%
  \BibitemOpen
  \bibfield  {author} {\bibinfo {author} {\bibfnamefont {R.~B.}\ \bibnamefont
  {Laughlin}},\ }\bibfield  {title} {\enquote {\bibinfo {title} {Quantized hall
  conductivity in two dimensions},}\ }\href {\doibase 10.1103/PhysRevB.23.5632}
  {\bibfield  {journal} {\bibinfo  {journal} {Phys. Rev. B}\ }\textbf {\bibinfo
  {volume} {23}},\ \bibinfo {pages} {5632--5633} (\bibinfo {year}
  {1981})}\BibitemShut {NoStop}%
\bibitem [{\citenamefont {Thouless}\ \emph {et~al.}(1982)\citenamefont
  {Thouless}, \citenamefont {Kohmoto}, \citenamefont {Nightingale},\ and\
  \citenamefont {den Nijs}}]{Thouless1982}%
  \BibitemOpen
  \bibfield  {author} {\bibinfo {author} {\bibfnamefont {D.~J.}\ \bibnamefont
  {Thouless}}, \bibinfo {author} {\bibfnamefont {M.}~\bibnamefont {Kohmoto}},
  \bibinfo {author} {\bibfnamefont {M.~P.}\ \bibnamefont {Nightingale}}, \ and\
  \bibinfo {author} {\bibfnamefont {M.}~\bibnamefont {den Nijs}},\ }\bibfield
  {title} {\enquote {\bibinfo {title} {Quantized hall conductance in a
  two-dimensional periodic potential},}\ }\href {\doibase
  10.1103/PhysRevLett.49.405} {\bibfield  {journal} {\bibinfo  {journal} {Phys.
  Rev. Lett.}\ }\textbf {\bibinfo {volume} {49}},\ \bibinfo {pages} {405--408}
  (\bibinfo {year} {1982})}\BibitemShut {NoStop}%
\bibitem [{\citenamefont {Ikebe}\ \emph {et~al.}(2010)\citenamefont {Ikebe},
  \citenamefont {Morimoto}, \citenamefont {Masutomi}, \citenamefont {Okamoto},
  \citenamefont {Aoki},\ and\ \citenamefont {Shimano}}]{Shimano2010}%
  \BibitemOpen
  \bibfield  {author} {\bibinfo {author} {\bibfnamefont {Y.}~\bibnamefont
  {Ikebe}}, \bibinfo {author} {\bibfnamefont {T.}~\bibnamefont {Morimoto}},
  \bibinfo {author} {\bibfnamefont {R.}~\bibnamefont {Masutomi}}, \bibinfo
  {author} {\bibfnamefont {T.}~\bibnamefont {Okamoto}}, \bibinfo {author}
  {\bibfnamefont {H.}~\bibnamefont {Aoki}}, \ and\ \bibinfo {author}
  {\bibfnamefont {R.}~\bibnamefont {Shimano}},\ }\bibfield  {title} {\enquote
  {\bibinfo {title} {Optical hall effect in the integer quantum hall regime},}\
  }\href {\doibase 10.1103/PhysRevLett.104.256802} {\bibfield  {journal}
  {\bibinfo  {journal} {Phys. Rev. Lett.}\ }\textbf {\bibinfo {volume} {104}},\
  \bibinfo {pages} {256802} (\bibinfo {year} {2010})}\BibitemShut {NoStop}%
\bibitem [{\citenamefont {Kane}\ and\ \citenamefont {Mele}(2005)}]{Kane2005}%
  \BibitemOpen
  \bibfield  {author} {\bibinfo {author} {\bibfnamefont {C.~L.}\ \bibnamefont
  {Kane}}\ and\ \bibinfo {author} {\bibfnamefont {E.~J.}\ \bibnamefont
  {Mele}},\ }\bibfield  {title} {\enquote {\bibinfo {title} {Quantum spin hall
  effect in graphene},}\ }\href {\doibase 10.1103/PhysRevLett.95.226801}
  {\bibfield  {journal} {\bibinfo  {journal} {Phys. Rev. Lett.}\ }\textbf
  {\bibinfo {volume} {95}},\ \bibinfo {pages} {226801} (\bibinfo {year}
  {2005})}\BibitemShut {NoStop}%
\bibitem [{\citenamefont {Bernevig}\ \emph {et~al.}(2006)\citenamefont
  {Bernevig}, \citenamefont {Hughes},\ and\ \citenamefont
  {Zhang}}]{Bernevig1757}%
  \BibitemOpen
  \bibfield  {author} {\bibinfo {author} {\bibfnamefont {B.~Andrei}\
  \bibnamefont {Bernevig}}, \bibinfo {author} {\bibfnamefont {Taylor~L.}\
  \bibnamefont {Hughes}}, \ and\ \bibinfo {author} {\bibfnamefont {Shou-Cheng}\
  \bibnamefont {Zhang}},\ }\bibfield  {title} {\enquote {\bibinfo {title}
  {Quantum spin hall effect and topological phase transition in hgte quantum
  wells},}\ }\href {\doibase 10.1126/science.1133734} {\bibfield  {journal}
  {\bibinfo  {journal} {Science}\ }\textbf {\bibinfo {volume} {314}},\ \bibinfo
  {pages} {1757--1761} (\bibinfo {year} {2006})}\BibitemShut {NoStop}%
\bibitem [{\citenamefont {Lan}\ \emph {et~al.}(2016)\citenamefont {Lan},
  \citenamefont {Kong},\ and\ \citenamefont {Wen}}]{Tian2016}%
  \BibitemOpen
  \bibfield  {author} {\bibinfo {author} {\bibfnamefont {Tian}\ \bibnamefont
  {Lan}}, \bibinfo {author} {\bibfnamefont {Liang}\ \bibnamefont {Kong}}, \
  and\ \bibinfo {author} {\bibfnamefont {Xiao-Gang}\ \bibnamefont {Wen}},\
  }\bibfield  {title} {\enquote {\bibinfo {title} {Theory of (2+1)-dimensional
  fermionic topological orders and fermionic/bosonic topological orders with
  symmetries},}\ }\href {\doibase 10.1103/PhysRevB.94.155113} {\bibfield
  {journal} {\bibinfo  {journal} {Phys. Rev. B}\ }\textbf {\bibinfo {volume}
  {94}},\ \bibinfo {pages} {155113} (\bibinfo {year} {2016})}\BibitemShut
  {NoStop}%
\bibitem [{\citenamefont {Chen}\ \emph {et~al.}(2012)\citenamefont {Chen},
  \citenamefont {Gu}, \citenamefont {Liu},\ and\ \citenamefont
  {Wen}}]{Chen1604}%
  \BibitemOpen
  \bibfield  {author} {\bibinfo {author} {\bibfnamefont {Xie}\ \bibnamefont
  {Chen}}, \bibinfo {author} {\bibfnamefont {Zheng-Cheng}\ \bibnamefont {Gu}},
  \bibinfo {author} {\bibfnamefont {Zheng-Xin}\ \bibnamefont {Liu}}, \ and\
  \bibinfo {author} {\bibfnamefont {Xiao-Gang}\ \bibnamefont {Wen}},\
  }\bibfield  {title} {\enquote {\bibinfo {title} {Symmetry-protected
  topological orders in interacting bosonic systems},}\ }\href {\doibase
  10.1126/science.1227224} {\bibfield  {journal} {\bibinfo  {journal}
  {Science}\ }\textbf {\bibinfo {volume} {338}},\ \bibinfo {pages} {1604--1606}
  (\bibinfo {year} {2012})}\BibitemShut {NoStop}%
\bibitem [{\citenamefont {Vishwanath}\ and\ \citenamefont
  {Senthil}(2013)}]{Vishwanath2013}%
  \BibitemOpen
  \bibfield  {author} {\bibinfo {author} {\bibfnamefont {Ashvin}\ \bibnamefont
  {Vishwanath}}\ and\ \bibinfo {author} {\bibfnamefont {T.}~\bibnamefont
  {Senthil}},\ }\bibfield  {title} {\enquote {\bibinfo {title} {Physics of
  three-dimensional bosonic topological insulators: Surface-deconfined
  criticality and quantized magnetoelectric effect},}\ }\href {\doibase
  10.1103/PhysRevX.3.011016} {\bibfield  {journal} {\bibinfo  {journal} {Phys.
  Rev. X}\ }\textbf {\bibinfo {volume} {3}},\ \bibinfo {pages} {011016}
  (\bibinfo {year} {2013})}\BibitemShut {NoStop}%
\bibitem [{\citenamefont {Metlitski}\ \emph {et~al.}(2013)\citenamefont
  {Metlitski}, \citenamefont {Kane},\ and\ \citenamefont
  {Fisher}}]{Metlitski2013}%
  \BibitemOpen
  \bibfield  {author} {\bibinfo {author} {\bibfnamefont {Max~A.}\ \bibnamefont
  {Metlitski}}, \bibinfo {author} {\bibfnamefont {C.~L.}\ \bibnamefont {Kane}},
  \ and\ \bibinfo {author} {\bibfnamefont {Matthew P.~A.}\ \bibnamefont
  {Fisher}},\ }\bibfield  {title} {\enquote {\bibinfo {title} {Bosonic
  topological insulator in three dimensions and the statistical witten
  effect},}\ }\href {\doibase 10.1103/PhysRevB.88.035131} {\bibfield  {journal}
  {\bibinfo  {journal} {Phys. Rev. B}\ }\textbf {\bibinfo {volume} {88}},\
  \bibinfo {pages} {035131} (\bibinfo {year} {2013})}\BibitemShut {NoStop}%
\bibitem [{\citenamefont {Senthil}\ and\ \citenamefont
  {Levin}(2013)}]{Senthil2013}%
  \BibitemOpen
  \bibfield  {author} {\bibinfo {author} {\bibfnamefont {T.}~\bibnamefont
  {Senthil}}\ and\ \bibinfo {author} {\bibfnamefont {Michael}\ \bibnamefont
  {Levin}},\ }\bibfield  {title} {\enquote {\bibinfo {title} {Integer quantum
  hall effect for bosons},}\ }\href {\doibase 10.1103/PhysRevLett.110.046801}
  {\bibfield  {journal} {\bibinfo  {journal} {Phys. Rev. Lett.}\ }\textbf
  {\bibinfo {volume} {110}},\ \bibinfo {pages} {046801} (\bibinfo {year}
  {2013})}\BibitemShut {NoStop}%
\bibitem [{\citenamefont {Lu}\ and\ \citenamefont
  {Vishwanath}(2012)}]{Vishwanath2012}%
  \BibitemOpen
  \bibfield  {author} {\bibinfo {author} {\bibfnamefont {Yuan-Ming}\
  \bibnamefont {Lu}}\ and\ \bibinfo {author} {\bibfnamefont {Ashvin}\
  \bibnamefont {Vishwanath}},\ }\bibfield  {title} {\enquote {\bibinfo {title}
  {Theory and classification of interacting integer topological phases in two
  dimensions: A chern-simons approach},}\ }\href {\doibase
  10.1103/PhysRevB.86.125119} {\bibfield  {journal} {\bibinfo  {journal} {Phys.
  Rev. B}\ }\textbf {\bibinfo {volume} {86}},\ \bibinfo {pages} {125119}
  (\bibinfo {year} {2012})}\BibitemShut {NoStop}%
\bibitem [{\citenamefont {Chen}\ \emph {et~al.}(2011)\citenamefont {Chen},
  \citenamefont {Liu},\ and\ \citenamefont {Wen}}]{Wen2011}%
  \BibitemOpen
  \bibfield  {author} {\bibinfo {author} {\bibfnamefont {Xie}\ \bibnamefont
  {Chen}}, \bibinfo {author} {\bibfnamefont {Zheng-Xin}\ \bibnamefont {Liu}}, \
  and\ \bibinfo {author} {\bibfnamefont {Xiao-Gang}\ \bibnamefont {Wen}},\
  }\bibfield  {title} {\enquote {\bibinfo {title} {Two-dimensional
  symmetry-protected topological orders and their protected gapless edge
  excitations},}\ }\href {\doibase 10.1103/PhysRevB.84.235141} {\bibfield
  {journal} {\bibinfo  {journal} {Phys. Rev. B}\ }\textbf {\bibinfo {volume}
  {84}},\ \bibinfo {pages} {235141} (\bibinfo {year} {2011})}\BibitemShut
  {NoStop}%
\bibitem [{\citenamefont {Van~Mechelen}\ and\ \citenamefont
  {Jacob}(2017)}]{van_mechelen2017}%
  \BibitemOpen
  \bibfield  {author} {\bibinfo {author} {\bibfnamefont {Todd}\ \bibnamefont
  {Van~Mechelen}}\ and\ \bibinfo {author} {\bibfnamefont {Zubin}\ \bibnamefont
  {Jacob}},\ }\bibfield  {title} {\enquote {\bibinfo {title} {Dirac-{Maxwell}
  correspondence: {Spin}-1 bosonic topological insulator},}\ }\href
  {http://arxiv.org/abs/1708.08192} {\bibfield  {journal} {\bibinfo  {journal}
  {arXiv: 1708.08192}\ } (\bibinfo {year} {2017})}\BibitemShut {NoStop}%
\bibitem [{\citenamefont {Hafezi}\ \emph {et~al.}(2013)\citenamefont {Hafezi},
  \citenamefont {Mittal}, \citenamefont {Fan}, \citenamefont {Migdall},\ and\
  \citenamefont {Taylor}}]{HafeziM.2013}%
  \BibitemOpen
  \bibfield  {author} {\bibinfo {author} {\bibfnamefont {M.}~\bibnamefont
  {Hafezi}}, \bibinfo {author} {\bibfnamefont {S.}~\bibnamefont {Mittal}},
  \bibinfo {author} {\bibfnamefont {J.}~\bibnamefont {Fan}}, \bibinfo {author}
  {\bibfnamefont {A.}~\bibnamefont {Migdall}}, \ and\ \bibinfo {author}
  {\bibfnamefont {J.~M.}\ \bibnamefont {Taylor}},\ }\bibfield  {title}
  {\enquote {\bibinfo {title} {Imaging topological edge states in silicon
  photonics},}\ }\href {http://dx.doi.org/10.1038/nphoton.2013.274} {\bibfield
  {journal} {\bibinfo  {journal} {Nat Photon}\ }\textbf {\bibinfo {volume}
  {7}},\ \bibinfo {pages} {1001--1005} (\bibinfo {year} {2013})},\ \bibinfo
  {note} {article}\BibitemShut {NoStop}%
\bibitem [{\citenamefont {Karzig}\ \emph {et~al.}(2015)\citenamefont {Karzig},
  \citenamefont {Bardyn}, \citenamefont {Lindner},\ and\ \citenamefont
  {Refael}}]{Karzig2015}%
  \BibitemOpen
  \bibfield  {author} {\bibinfo {author} {\bibfnamefont {Torsten}\ \bibnamefont
  {Karzig}}, \bibinfo {author} {\bibfnamefont {Charles-Edouard}\ \bibnamefont
  {Bardyn}}, \bibinfo {author} {\bibfnamefont {Netanel~H.}\ \bibnamefont
  {Lindner}}, \ and\ \bibinfo {author} {\bibfnamefont {Gil}\ \bibnamefont
  {Refael}},\ }\bibfield  {title} {\enquote {\bibinfo {title} {Topological
  polaritons},}\ }\href {\doibase 10.1103/PhysRevX.5.031001} {\bibfield
  {journal} {\bibinfo  {journal} {Phys. Rev. X}\ }\textbf {\bibinfo {volume}
  {5}},\ \bibinfo {pages} {031001} (\bibinfo {year} {2015})}\BibitemShut
  {NoStop}%
\bibitem [{\citenamefont {Hadad}\ \emph {et~al.}(2017)\citenamefont {Hadad},
  \citenamefont {Vitelli},\ and\ \citenamefont {Alu}}]{Alu2017}%
  \BibitemOpen
  \bibfield  {author} {\bibinfo {author} {\bibfnamefont {Yakir}\ \bibnamefont
  {Hadad}}, \bibinfo {author} {\bibfnamefont {Vincenzo}\ \bibnamefont
  {Vitelli}}, \ and\ \bibinfo {author} {\bibfnamefont {Andrea}\ \bibnamefont
  {Alu}},\ }\bibfield  {title} {\enquote {\bibinfo {title} {Solitons and
  propagating domain walls in topological resonator arrays},}\ }\href {\doibase
  10.1021/acsphotonics.7b00303} {\bibfield  {journal} {\bibinfo  {journal} {ACS
  Photonics}\ }\textbf {\bibinfo {volume} {4}},\ \bibinfo {pages} {1974--1979}
  (\bibinfo {year} {2017})}\BibitemShut {NoStop}%
\bibitem [{\citenamefont {De~Nittis}\ and\ \citenamefont
  {Lein}(2017)}]{Lein2017}%
  \BibitemOpen
  \bibfield  {author} {\bibinfo {author} {\bibfnamefont {Giuseppe}\
  \bibnamefont {De~Nittis}}\ and\ \bibinfo {author} {\bibfnamefont {Max}\
  \bibnamefont {Lein}},\ }\bibfield  {title} {\enquote {\bibinfo {title}
  {Symmetry {Classification} of {Topological} {Photonic} {Crystals}},}\ }\href
  {http://arxiv.org/abs/1710.08104} {\bibfield  {journal} {\bibinfo  {journal}
  {arXiv:1710.08104}\ } (\bibinfo {year} {2017})}\BibitemShut {NoStop}%
\bibitem [{\citenamefont {Gu}\ \emph {et~al.}(2015)\citenamefont {Gu},
  \citenamefont {Wang},\ and\ \citenamefont {Wen}}]{Zheng2015}%
  \BibitemOpen
  \bibfield  {author} {\bibinfo {author} {\bibfnamefont {Zheng-Cheng}\
  \bibnamefont {Gu}}, \bibinfo {author} {\bibfnamefont {Zhenghan}\ \bibnamefont
  {Wang}}, \ and\ \bibinfo {author} {\bibfnamefont {Xiao-Gang}\ \bibnamefont
  {Wen}},\ }\bibfield  {title} {\enquote {\bibinfo {title} {Classification of
  two-dimensional fermionic and bosonic topological orders},}\ }\href {\doibase
  10.1103/PhysRevB.91.125149} {\bibfield  {journal} {\bibinfo  {journal} {Phys.
  Rev. B}\ }\textbf {\bibinfo {volume} {91}},\ \bibinfo {pages} {125149}
  (\bibinfo {year} {2015})}\BibitemShut {NoStop}%
\bibitem [{\citenamefont {Lu}\ \emph {et~al.}(2013)\citenamefont {Lu},
  \citenamefont {Fu}, \citenamefont {Joannopoulos},\ and\ \citenamefont
  {Soljacic}}]{Lu2013}%
  \BibitemOpen
  \bibfield  {author} {\bibinfo {author} {\bibfnamefont {Ling}\ \bibnamefont
  {Lu}}, \bibinfo {author} {\bibfnamefont {Liang}\ \bibnamefont {Fu}}, \bibinfo
  {author} {\bibfnamefont {John~D.}\ \bibnamefont {Joannopoulos}}, \ and\
  \bibinfo {author} {\bibfnamefont {Marin}\ \bibnamefont {Soljacic}},\
  }\bibfield  {title} {\enquote {\bibinfo {title} {Weyl points and line nodes
  in gyroid photonic crystals},}\ }\href {\doibase 10.1038/nphoton.2013.42}
  {\bibfield  {journal} {\bibinfo  {journal} {Nat Photon}\ }\textbf {\bibinfo
  {volume} {7}},\ \bibinfo {pages} {294--299} (\bibinfo {year}
  {2013})}\BibitemShut {NoStop}%
\bibitem [{\citenamefont {Wang}\ \emph {et~al.}(2009)\citenamefont {Wang},
  \citenamefont {Chong}, \citenamefont {Joannopoulos},\ and\ \citenamefont
  {Soljacic}}]{Wang2009}%
  \BibitemOpen
  \bibfield  {author} {\bibinfo {author} {\bibfnamefont {Zheng}\ \bibnamefont
  {Wang}}, \bibinfo {author} {\bibfnamefont {Yidong}\ \bibnamefont {Chong}},
  \bibinfo {author} {\bibfnamefont {J.~D.}\ \bibnamefont {Joannopoulos}}, \
  and\ \bibinfo {author} {\bibfnamefont {Marin}\ \bibnamefont {Soljacic}},\
  }\bibfield  {title} {\enquote {\bibinfo {title} {Observation of
  unidirectional backscattering-immune topological electromagnetic states},}\
  }\href {\doibase 10.1038/nature08293} {\bibfield  {journal} {\bibinfo
  {journal} {Nature}\ }\textbf {\bibinfo {volume} {461}},\ \bibinfo {pages}
  {772--775} (\bibinfo {year} {2009})}\BibitemShut {NoStop}%
\bibitem [{\citenamefont {Wang}\ and\ \citenamefont {Fan}(2005)}]{Wang:05}%
  \BibitemOpen
  \bibfield  {author} {\bibinfo {author} {\bibfnamefont {Zheng}\ \bibnamefont
  {Wang}}\ and\ \bibinfo {author} {\bibfnamefont {Shanhui}\ \bibnamefont
  {Fan}},\ }\bibfield  {title} {\enquote {\bibinfo {title} {Optical circulators
  in two-dimensional magneto-optical photonic crystals},}\ }\href {\doibase
  10.1364/OL.30.001989} {\bibfield  {journal} {\bibinfo  {journal} {Opt.
  Lett.}\ }\textbf {\bibinfo {volume} {30}},\ \bibinfo {pages} {1989--1991}
  (\bibinfo {year} {2005})}\BibitemShut {NoStop}%
\bibitem [{\citenamefont {Rechtsman}\ \emph {et~al.}(2013)\citenamefont
  {Rechtsman}, \citenamefont {Zeuner}, \citenamefont {Plotnik}, \citenamefont
  {Lumer}, \citenamefont {Podolsky}, \citenamefont {Dreisow}, \citenamefont
  {Nolte}, \citenamefont {Segev},\ and\ \citenamefont
  {Szameit}}]{Rechtsman2013}%
  \BibitemOpen
  \bibfield  {author} {\bibinfo {author} {\bibfnamefont {Mikael~C.}\
  \bibnamefont {Rechtsman}}, \bibinfo {author} {\bibfnamefont {Julia~M.}\
  \bibnamefont {Zeuner}}, \bibinfo {author} {\bibfnamefont {Yonatan}\
  \bibnamefont {Plotnik}}, \bibinfo {author} {\bibfnamefont {Yaakov}\
  \bibnamefont {Lumer}}, \bibinfo {author} {\bibfnamefont {Daniel}\
  \bibnamefont {Podolsky}}, \bibinfo {author} {\bibfnamefont {Felix}\
  \bibnamefont {Dreisow}}, \bibinfo {author} {\bibfnamefont {Stefan}\
  \bibnamefont {Nolte}}, \bibinfo {author} {\bibfnamefont {Mordechai}\
  \bibnamefont {Segev}}, \ and\ \bibinfo {author} {\bibfnamefont {Alexander}\
  \bibnamefont {Szameit}},\ }\bibfield  {title} {\enquote {\bibinfo {title}
  {Photonic floquet topological insulators},}\ }\href
  {http://dx.doi.org/10.1038/nature12066} {\bibfield  {journal} {\bibinfo
  {journal} {Nature}\ }\textbf {\bibinfo {volume} {496}},\ \bibinfo {pages}
  {196--200} (\bibinfo {year} {2013})},\ \bibinfo {note} {letter}\BibitemShut
  {NoStop}%
\bibitem [{\citenamefont {Khanikaev}\ \emph {et~al.}(2013)\citenamefont
  {Khanikaev}, \citenamefont {Hossein~Mousavi}, \citenamefont {Tse},
  \citenamefont {Kargarian}, \citenamefont {MacDonald},\ and\ \citenamefont
  {Shvets}}]{Khanikaev2013}%
  \BibitemOpen
  \bibfield  {author} {\bibinfo {author} {\bibfnamefont {Alexander~B.}\
  \bibnamefont {Khanikaev}}, \bibinfo {author} {\bibfnamefont {S.}~\bibnamefont
  {Hossein~Mousavi}}, \bibinfo {author} {\bibfnamefont {Wang-Kong}\
  \bibnamefont {Tse}}, \bibinfo {author} {\bibfnamefont {Mehdi}\ \bibnamefont
  {Kargarian}}, \bibinfo {author} {\bibfnamefont {Allan~H.}\ \bibnamefont
  {MacDonald}}, \ and\ \bibinfo {author} {\bibfnamefont {Gennady}\ \bibnamefont
  {Shvets}},\ }\bibfield  {title} {\enquote {\bibinfo {title} {Photonic
  topological insulators},}\ }\href {\doibase 10.1038/nmat3520} {\bibfield
  {journal} {\bibinfo  {journal} {Nat Mater}\ }\textbf {\bibinfo {volume}
  {12}},\ \bibinfo {pages} {233--239} (\bibinfo {year} {2013})}\BibitemShut
  {NoStop}%
\bibitem [{\citenamefont {Slobozhanyuk}\ \emph {et~al.}(2017)\citenamefont
  {Slobozhanyuk}, \citenamefont {Mousavi}, \citenamefont {Ni}, \citenamefont
  {Smirnova}, \citenamefont {Kivshar},\ and\ \citenamefont
  {Khanikaev}}]{Slobozhanyuk2017}%
  \BibitemOpen
  \bibfield  {author} {\bibinfo {author} {\bibfnamefont {Alexey}\ \bibnamefont
  {Slobozhanyuk}}, \bibinfo {author} {\bibfnamefont {S.~Hossein}\ \bibnamefont
  {Mousavi}}, \bibinfo {author} {\bibfnamefont {Xiang}\ \bibnamefont {Ni}},
  \bibinfo {author} {\bibfnamefont {Daria}\ \bibnamefont {Smirnova}}, \bibinfo
  {author} {\bibfnamefont {Yuri~S.}\ \bibnamefont {Kivshar}}, \ and\ \bibinfo
  {author} {\bibfnamefont {Alexander~B.}\ \bibnamefont {Khanikaev}},\
  }\bibfield  {title} {\enquote {\bibinfo {title} {Three-dimensional
  all-dielectric photonic topological insulator},}\ }\href
  {http://dx.doi.org/10.1038/nphoton.2016.253} {\bibfield  {journal} {\bibinfo
  {journal} {Nat Photon}\ }\textbf {\bibinfo {volume} {11}},\ \bibinfo {pages}
  {130--136} (\bibinfo {year} {2017})},\ \bibinfo {note} {article}\BibitemShut
  {NoStop}%
\bibitem [{\citenamefont {He}\ \emph {et~al.}(2016)\citenamefont {He},
  \citenamefont {Sun}, \citenamefont {Liu}, \citenamefont {Lu}, \citenamefont
  {Chen}, \citenamefont {Feng},\ and\ \citenamefont {Chen}}]{He4924}%
  \BibitemOpen
  \bibfield  {author} {\bibinfo {author} {\bibfnamefont {Cheng}\ \bibnamefont
  {He}}, \bibinfo {author} {\bibfnamefont {Xiao-Chen}\ \bibnamefont {Sun}},
  \bibinfo {author} {\bibfnamefont {Xiao-Ping}\ \bibnamefont {Liu}}, \bibinfo
  {author} {\bibfnamefont {Ming-Hui}\ \bibnamefont {Lu}}, \bibinfo {author}
  {\bibfnamefont {Yulin}\ \bibnamefont {Chen}}, \bibinfo {author}
  {\bibfnamefont {Liang}\ \bibnamefont {Feng}}, \ and\ \bibinfo {author}
  {\bibfnamefont {Yan-Feng}\ \bibnamefont {Chen}},\ }\bibfield  {title}
  {\enquote {\bibinfo {title} {Photonic topological insulator with broken
  time-reversal symmetry},}\ }\href {\doibase 10.1073/pnas.1525502113}
  {\bibfield  {journal} {\bibinfo  {journal} {Proceedings of the National
  Academy of Sciences}\ }\textbf {\bibinfo {volume} {113}},\ \bibinfo {pages}
  {4924--4928} (\bibinfo {year} {2016})}\BibitemShut {NoStop}%
\bibitem [{\citenamefont {Raghu}\ and\ \citenamefont
  {Haldane}(2008)}]{Haldane2008}%
  \BibitemOpen
  \bibfield  {author} {\bibinfo {author} {\bibfnamefont {S.}~\bibnamefont
  {Raghu}}\ and\ \bibinfo {author} {\bibfnamefont {F.~D.~M.}\ \bibnamefont
  {Haldane}},\ }\bibfield  {title} {\enquote {\bibinfo {title} {Analogs of
  quantum-hall-effect edge states in photonic crystals},}\ }\href {\doibase
  10.1103/PhysRevA.78.033834} {\bibfield  {journal} {\bibinfo  {journal} {Phys.
  Rev. A}\ }\textbf {\bibinfo {volume} {78}},\ \bibinfo {pages} {033834}
  (\bibinfo {year} {2008})}\BibitemShut {NoStop}%
\bibitem [{\citenamefont {Koch}\ \emph {et~al.}(2010)\citenamefont {Koch},
  \citenamefont {Houck}, \citenamefont {Hur},\ and\ \citenamefont
  {Girvin}}]{Houck2010}%
  \BibitemOpen
  \bibfield  {author} {\bibinfo {author} {\bibfnamefont {Jens}\ \bibnamefont
  {Koch}}, \bibinfo {author} {\bibfnamefont {Andrew~A.}\ \bibnamefont {Houck}},
  \bibinfo {author} {\bibfnamefont {Karyn~Le}\ \bibnamefont {Hur}}, \ and\
  \bibinfo {author} {\bibfnamefont {S.~M.}\ \bibnamefont {Girvin}},\ }\bibfield
   {title} {\enquote {\bibinfo {title} {Time-reversal-symmetry breaking in
  circuit-qed-based photon lattices},}\ }\href {\doibase
  10.1103/PhysRevA.82.043811} {\bibfield  {journal} {\bibinfo  {journal} {Phys.
  Rev. A}\ }\textbf {\bibinfo {volume} {82}},\ \bibinfo {pages} {043811}
  (\bibinfo {year} {2010})}\BibitemShut {NoStop}%
\bibitem [{\citenamefont {Jaksch}\ and\ \citenamefont
  {Zoller}(2003)}]{Zoller2003}%
  \BibitemOpen
  \bibfield  {author} {\bibinfo {author} {\bibfnamefont {D}~\bibnamefont
  {Jaksch}}\ and\ \bibinfo {author} {\bibfnamefont {P}~\bibnamefont {Zoller}},\
  }\bibfield  {title} {\enquote {\bibinfo {title} {Creation of effective
  magnetic fields in optical lattices: the hofstadter butterfly for cold
  neutral atoms},}\ }\href {http://stacks.iop.org/1367-2630/5/i=1/a=356}
  {\bibfield  {journal} {\bibinfo  {journal} {New Journal of Physics}\ }\textbf
  {\bibinfo {volume} {5}},\ \bibinfo {pages} {56} (\bibinfo {year}
  {2003})}\BibitemShut {NoStop}%
\bibitem [{\citenamefont {Silveirinha}(2015)}]{Silveirinha2015}%
  \BibitemOpen
  \bibfield  {author} {\bibinfo {author} {\bibfnamefont {M\'ario~G.}\
  \bibnamefont {Silveirinha}},\ }\bibfield  {title} {\enquote {\bibinfo {title}
  {Chern invariants for continuous media},}\ }\href {\doibase
  10.1103/PhysRevB.92.125153} {\bibfield  {journal} {\bibinfo  {journal} {Phys.
  Rev. B}\ }\textbf {\bibinfo {volume} {92}},\ \bibinfo {pages} {125153}
  (\bibinfo {year} {2015})}\BibitemShut {NoStop}%
\bibitem [{\citenamefont {Silveirinha}(2017)}]{Silveirinha2017}%
  \BibitemOpen
  \bibfield  {author} {\bibinfo {author} {\bibfnamefont {M\'ario~G.}\
  \bibnamefont {Silveirinha}},\ }\bibfield  {title} {\enquote {\bibinfo {title}
  {$\mathcal{P}\ifmmode\cdot\else\textperiodcentered\fi{}\mathcal{T}\ifmmode\cdot\else\textperiodcentered\fi{}\mathcal{D}$
  symmetry-protected scattering anomaly in optics},}\ }\href {\doibase
  10.1103/PhysRevB.95.035153} {\bibfield  {journal} {\bibinfo  {journal} {Phys.
  Rev. B}\ }\textbf {\bibinfo {volume} {95}},\ \bibinfo {pages} {035153}
  (\bibinfo {year} {2017})}\BibitemShut {NoStop}%
\bibitem [{\citenamefont {Jin}\ \emph {et~al.}(2016)\citenamefont {Jin},
  \citenamefont {Lu}, \citenamefont {Wang}, \citenamefont {Fang}, \citenamefont
  {Joannopoulos}, \citenamefont {Soljacic}, \citenamefont {Fu},\ and\
  \citenamefont {Fang}}]{Jin2016}%
  \BibitemOpen
  \bibfield  {author} {\bibinfo {author} {\bibfnamefont {Dafei}\ \bibnamefont
  {Jin}}, \bibinfo {author} {\bibfnamefont {Ling}\ \bibnamefont {Lu}}, \bibinfo
  {author} {\bibfnamefont {Zhong}\ \bibnamefont {Wang}}, \bibinfo {author}
  {\bibfnamefont {Chen}\ \bibnamefont {Fang}}, \bibinfo {author} {\bibfnamefont
  {John~D.}\ \bibnamefont {Joannopoulos}}, \bibinfo {author} {\bibfnamefont
  {Marin}\ \bibnamefont {Soljacic}}, \bibinfo {author} {\bibfnamefont {Liang}\
  \bibnamefont {Fu}}, \ and\ \bibinfo {author} {\bibfnamefont {Nicholas~X.}\
  \bibnamefont {Fang}},\ }\bibfield  {title} {\enquote {\bibinfo {title}
  {Topological magnetoplasmon},}\ }\href
  {http://dx.doi.org/10.1038/ncomms13486} {\bibfield  {journal} {\bibinfo
  {journal} {Nature Communications}\ }\textbf {\bibinfo {volume} {7}},\
  \bibinfo {pages} {13486 EP --} (\bibinfo {year} {2016})},\ \bibinfo {note}
  {article}\BibitemShut {NoStop}%
\bibitem [{\citenamefont {Shi}\ and\ \citenamefont {Song}(2018)}]{Song2018}%
  \BibitemOpen
  \bibfield  {author} {\bibinfo {author} {\bibfnamefont {Li-kun}\ \bibnamefont
  {Shi}}\ and\ \bibinfo {author} {\bibfnamefont {Justin C.~W.}\ \bibnamefont
  {Song}},\ }\bibfield  {title} {\enquote {\bibinfo {title} {Plasmon geometric
  phase and plasmon hall shift},}\ }\href {\doibase 10.1103/PhysRevX.8.021020}
  {\bibfield  {journal} {\bibinfo  {journal} {Phys. Rev. X}\ }\textbf {\bibinfo
  {volume} {8}},\ \bibinfo {pages} {021020} (\bibinfo {year}
  {2018})}\BibitemShut {NoStop}%
\bibitem [{\citenamefont {Bialynicki-Birula}\ and\ \citenamefont
  {Bialynicka-Birula}(1987)}]{BB1987}%
  \BibitemOpen
  \bibfield  {author} {\bibinfo {author} {\bibfnamefont {Iwo}\ \bibnamefont
  {Bialynicki-Birula}}\ and\ \bibinfo {author} {\bibfnamefont {Zofia}\
  \bibnamefont {Bialynicka-Birula}},\ }\bibfield  {title} {\enquote {\bibinfo
  {title} {Berry's phase in the relativistic theory of spinning particles},}\
  }\href {\doibase 10.1103/PhysRevD.35.2383} {\bibfield  {journal} {\bibinfo
  {journal} {Phys. Rev. D}\ }\textbf {\bibinfo {volume} {35}},\ \bibinfo
  {pages} {2383--2387} (\bibinfo {year} {1987})}\BibitemShut {NoStop}%
\bibitem [{\citenamefont {Stone}(2016)}]{Stone2016}%
  \BibitemOpen
  \bibfield  {author} {\bibinfo {author} {\bibfnamefont {Michael}\ \bibnamefont
  {Stone}},\ }\bibfield  {title} {\enquote {\bibinfo {title} {Berry phase and
  anomalous velocity of weyl fermions and maxwell photons},}\ }\href {\doibase
  10.1142/S0217979215502495} {\bibfield  {journal} {\bibinfo  {journal}
  {International Journal of Modern Physics B}\ }\textbf {\bibinfo {volume}
  {30}},\ \bibinfo {pages} {1550249} (\bibinfo {year} {2016})}\BibitemShut
  {NoStop}%
\bibitem [{\citenamefont {Hasan}\ and\ \citenamefont {Kane}(2010)}]{Hasan2010}%
  \BibitemOpen
  \bibfield  {author} {\bibinfo {author} {\bibfnamefont {M.~Z.}\ \bibnamefont
  {Hasan}}\ and\ \bibinfo {author} {\bibfnamefont {C.~L.}\ \bibnamefont
  {Kane}},\ }\bibfield  {title} {\enquote {\bibinfo {title} {Colloquium:
  Topological insulators},}\ }\href {\doibase 10.1103/RevModPhys.82.3045}
  {\bibfield  {journal} {\bibinfo  {journal} {Rev. Mod. Phys.}\ }\textbf
  {\bibinfo {volume} {82}},\ \bibinfo {pages} {3045--3067} (\bibinfo {year}
  {2010})}\BibitemShut {NoStop}%
\bibitem [{\citenamefont {Barnett}\ \emph {et~al.}(2016)\citenamefont
  {Barnett}, \citenamefont {Allen}, \citenamefont {Cameron}, \citenamefont
  {Gilson}, \citenamefont {Padgett}, \citenamefont {Speirits},\ and\
  \citenamefont {Yao}}]{barnett_natures_2016}%
  \BibitemOpen
  \bibfield  {author} {\bibinfo {author} {\bibfnamefont {Stephen~M.}\
  \bibnamefont {Barnett}}, \bibinfo {author} {\bibfnamefont {L.}~\bibnamefont
  {Allen}}, \bibinfo {author} {\bibfnamefont {Robert~P.}\ \bibnamefont
  {Cameron}}, \bibinfo {author} {\bibfnamefont {Claire~R.}\ \bibnamefont
  {Gilson}}, \bibinfo {author} {\bibfnamefont {Miles~J.}\ \bibnamefont
  {Padgett}}, \bibinfo {author} {\bibfnamefont {Fiona~C.}\ \bibnamefont
  {Speirits}}, \ and\ \bibinfo {author} {\bibfnamefont {Alison~M.}\
  \bibnamefont {Yao}},\ }\bibfield  {title} {\enquote {\bibinfo {title} {On the
  natures of the spin and orbital parts of optical angular momentum},}\ }\href
  {\doibase 10.1088/2040-8978/18/6/064004} {\bibfield  {journal} {\bibinfo
  {journal} {Journal of Optics}\ }\textbf {\bibinfo {volume} {18}},\ \bibinfo
  {pages} {064004} (\bibinfo {year} {2016})}\BibitemShut {NoStop}%
\bibitem [{\citenamefont {Bliokh}\ and\ \citenamefont
  {Nori}(2015)}]{BLIOKH20151}%
  \BibitemOpen
  \bibfield  {author} {\bibinfo {author} {\bibfnamefont {Konstantin~Y.}\
  \bibnamefont {Bliokh}}\ and\ \bibinfo {author} {\bibfnamefont {Franco}\
  \bibnamefont {Nori}},\ }\bibfield  {title} {\enquote {\bibinfo {title}
  {Transverse and longitudinal angular momenta of light},}\ }\href {\doibase
  https://doi.org/10.1016/j.physrep.2015.06.003} {\bibfield  {journal}
  {\bibinfo  {journal} {Physics Reports}\ }\textbf {\bibinfo {volume} {592}},\
  \bibinfo {pages} {1 -- 38} (\bibinfo {year} {2015})}\BibitemShut {NoStop}%
\bibitem [{\citenamefont {Mechelen}\ and\ \citenamefont
  {Jacob}(2016)}]{VanMechelen:16}%
  \BibitemOpen
  \bibfield  {author} {\bibinfo {author} {\bibfnamefont {Todd~Van}\
  \bibnamefont {Mechelen}}\ and\ \bibinfo {author} {\bibfnamefont {Zubin}\
  \bibnamefont {Jacob}},\ }\bibfield  {title} {\enquote {\bibinfo {title}
  {Universal spin-momentum locking of evanescent waves},}\ }\href {\doibase
  10.1364/OPTICA.3.000118} {\bibfield  {journal} {\bibinfo  {journal} {Optica}\
  }\textbf {\bibinfo {volume} {3}},\ \bibinfo {pages} {118--126} (\bibinfo
  {year} {2016})}\BibitemShut {NoStop}%
\bibitem [{\citenamefont {Kalhor}\ \emph {et~al.}(2016)\citenamefont {Kalhor},
  \citenamefont {Thundat},\ and\ \citenamefont
  {Jacob}}]{kalhor_universal_2016}%
  \BibitemOpen
  \bibfield  {author} {\bibinfo {author} {\bibfnamefont {Farid}\ \bibnamefont
  {Kalhor}}, \bibinfo {author} {\bibfnamefont {Thomas}\ \bibnamefont
  {Thundat}}, \ and\ \bibinfo {author} {\bibfnamefont {Zubin}\ \bibnamefont
  {Jacob}},\ }\bibfield  {title} {\enquote {\bibinfo {title} {Universal
  spin-momentum locked optical forces},}\ }\href {\doibase 10.1063/1.4941539}
  {\bibfield  {journal} {\bibinfo  {journal} {Applied Physics Letters}\
  }\textbf {\bibinfo {volume} {108}},\ \bibinfo {pages} {061102} (\bibinfo
  {year} {2016})}\BibitemShut {NoStop}%
\bibitem [{\citenamefont {Dunne}(1999)}]{Dunne1999}%
  \BibitemOpen
  \bibfield  {author} {\bibinfo {author} {\bibfnamefont {G.~V.}\ \bibnamefont
  {Dunne}},\ }\bibfield  {title} {\enquote {\bibinfo {title} {Aspects of
  chern-simons theory},}\ }in\ \href@noop {} {\emph {\bibinfo {booktitle}
  {Aspects topologiques de la physique en basse dimension. Topological aspects
  of low dimensional systems}}},\ \bibinfo {editor} {edited by\ \bibinfo
  {editor} {\bibfnamefont {A.}~\bibnamefont {Comtet}}, \bibinfo {editor}
  {\bibfnamefont {T.}~\bibnamefont {Jolic{\oe}ur}}, \bibinfo {editor}
  {\bibfnamefont {S.}~\bibnamefont {Ouvry}}, \ and\ \bibinfo {editor}
  {\bibfnamefont {F.}~\bibnamefont {David}}}\ (\bibinfo  {publisher} {Springer
  Berlin Heidelberg},\ \bibinfo {address} {Berlin, Heidelberg},\ \bibinfo
  {year} {1999})\ pp.\ \bibinfo {pages} {177--263}\BibitemShut {NoStop}%
\bibitem [{\citenamefont {Boyanovsky}\ \emph {et~al.}(1986)\citenamefont
  {Boyanovsky}, \citenamefont {Blankenbecler},\ and\ \citenamefont
  {Yahalom}}]{BOYANOVSKY1986483}%
  \BibitemOpen
  \bibfield  {author} {\bibinfo {author} {\bibfnamefont {D.}~\bibnamefont
  {Boyanovsky}}, \bibinfo {author} {\bibfnamefont {R.}~\bibnamefont
  {Blankenbecler}}, \ and\ \bibinfo {author} {\bibfnamefont {R.}~\bibnamefont
  {Yahalom}},\ }\bibfield  {title} {\enquote {\bibinfo {title} {Physical origin
  of topological mass in 2 + 1 dimensions},}\ }\href {\doibase
  https://doi.org/10.1016/0550-3213(86)90564-X} {\bibfield  {journal} {\bibinfo
   {journal} {Nuclear Physics B}\ }\textbf {\bibinfo {volume} {270}},\ \bibinfo
  {pages} {483 -- 505} (\bibinfo {year} {1986})}\BibitemShut {NoStop}%
\bibitem [{\citenamefont {Ziegler}(1998)}]{Ziegler1998}%
  \BibitemOpen
  \bibfield  {author} {\bibinfo {author} {\bibfnamefont {K.}~\bibnamefont
  {Ziegler}},\ }\bibfield  {title} {\enquote {\bibinfo {title} {Delocalization
  of 2d dirac fermions: The role of a broken supersymmetry},}\ }\href {\doibase
  10.1103/PhysRevLett.80.3113} {\bibfield  {journal} {\bibinfo  {journal}
  {Phys. Rev. Lett.}\ }\textbf {\bibinfo {volume} {80}},\ \bibinfo {pages}
  {3113--3116} (\bibinfo {year} {1998})}\BibitemShut {NoStop}%
\bibitem [{\citenamefont {Gates~Jr}\ \emph {et~al.}(2001)\citenamefont
  {Gates~Jr}, \citenamefont {Grisaru}, \citenamefont {Rocek},\ and\
  \citenamefont {Siegel}}]{gates2001}%
  \BibitemOpen
  \bibfield  {author} {\bibinfo {author} {\bibfnamefont {S.~J.}\ \bibnamefont
  {Gates~Jr}}, \bibinfo {author} {\bibfnamefont {M.~T.}\ \bibnamefont
  {Grisaru}}, \bibinfo {author} {\bibfnamefont {M.}~\bibnamefont {Rocek}}, \
  and\ \bibinfo {author} {\bibfnamefont {W.}~\bibnamefont {Siegel}},\
  }\bibfield  {title} {\enquote {\bibinfo {title} {Superspace, or {One}
  thousand and one lessons in supersymmetry},}\ }\href
  {http://arxiv.org/abs/hep-th/0108200} {\bibfield  {journal} {\bibinfo
  {journal} {arXiv:hep-th/0108200}\ } (\bibinfo {year} {2001})}\BibitemShut
  {NoStop}%
\bibitem [{\citenamefont {Landau}\ \emph {et~al.}(2013)\citenamefont {Landau},
  \citenamefont {Bell}, \citenamefont {Kearsley}, \citenamefont {Pitaevskii},
  \citenamefont {Lifshitz},\ and\ \citenamefont
  {Sykes}}]{landau2013electrodynamics}%
  \BibitemOpen
  \bibfield  {author} {\bibinfo {author} {\bibfnamefont {Lev~Davidovich}\
  \bibnamefont {Landau}}, \bibinfo {author} {\bibfnamefont {JS}~\bibnamefont
  {Bell}}, \bibinfo {author} {\bibfnamefont {MJ}~\bibnamefont {Kearsley}},
  \bibinfo {author} {\bibfnamefont {LP}~\bibnamefont {Pitaevskii}}, \bibinfo
  {author} {\bibfnamefont {EM}~\bibnamefont {Lifshitz}}, \ and\ \bibinfo
  {author} {\bibfnamefont {JB}~\bibnamefont {Sykes}},\ }\href@noop {} {\emph
  {\bibinfo {title} {Electrodynamics of continuous media}}},\ Vol.~\bibinfo
  {volume} {8}\ (\bibinfo  {publisher} {elsevier},\ \bibinfo {year}
  {2013})\BibitemShut {NoStop}%
\bibitem [{\citenamefont {Horsley}\ and\ \citenamefont
  {Philbin}(2014)}]{Philbin2014}%
  \BibitemOpen
  \bibfield  {author} {\bibinfo {author} {\bibfnamefont {S~A~R}\ \bibnamefont
  {Horsley}}\ and\ \bibinfo {author} {\bibfnamefont {T~G}\ \bibnamefont
  {Philbin}},\ }\bibfield  {title} {\enquote {\bibinfo {title} {Canonical
  quantization of electromagnetism in spatially dispersive media},}\ }\href
  {http://stacks.iop.org/1367-2630/16/i=1/a=013030} {\bibfield  {journal}
  {\bibinfo  {journal} {New Journal of Physics}\ }\textbf {\bibinfo {volume}
  {16}},\ \bibinfo {pages} {013030} (\bibinfo {year} {2014})}\BibitemShut
  {NoStop}%
\bibitem [{\citenamefont {Ivanov}\ and\ \citenamefont
  {Nikolaev}(1996)}]{Nikolaev1996}%
  \BibitemOpen
  \bibfield  {author} {\bibinfo {author} {\bibfnamefont {Stefan~T}\
  \bibnamefont {Ivanov}}\ and\ \bibinfo {author} {\bibfnamefont {Nikolay~I}\
  \bibnamefont {Nikolaev}},\ }\bibfield  {title} {\enquote {\bibinfo {title}
  {Singular waves along the boundary of gyrotropic plasma},}\ }\href
  {http://stacks.iop.org/0022-3727/29/i=4/a=027} {\bibfield  {journal}
  {\bibinfo  {journal} {Journal of Physics D: Applied Physics}\ }\textbf
  {\bibinfo {volume} {29}},\ \bibinfo {pages} {1107} (\bibinfo {year}
  {1996})}\BibitemShut {NoStop}%
\bibitem [{\citenamefont {Eroglu}(2010)}]{eroglu2010wave}%
  \BibitemOpen
  \bibfield  {author} {\bibinfo {author} {\bibfnamefont {Abdullah}\
  \bibnamefont {Eroglu}},\ }\href@noop {} {\emph {\bibinfo {title} {Wave
  propagation and radiation in gyrotropic and anisotropic media}}}\ (\bibinfo
  {publisher} {Springer Science \& Business Media},\ \bibinfo {year}
  {2010})\BibitemShut {NoStop}%
\bibitem [{\citenamefont {Shan}\ \emph {et~al.}(2010)\citenamefont {Shan},
  \citenamefont {Lu},\ and\ \citenamefont {Shen}}]{Shen2010}%
  \BibitemOpen
  \bibfield  {author} {\bibinfo {author} {\bibfnamefont {Wen-Yu}\ \bibnamefont
  {Shan}}, \bibinfo {author} {\bibfnamefont {Hai-Zhou}\ \bibnamefont {Lu}}, \
  and\ \bibinfo {author} {\bibfnamefont {Shun-Qing}\ \bibnamefont {Shen}},\
  }\bibfield  {title} {\enquote {\bibinfo {title} {Effective continuous model
  for surface states and thin films of three-dimensional topological
  insulators},}\ }\href {http://stacks.iop.org/1367-2630/12/i=4/a=043048}
  {\bibfield  {journal} {\bibinfo  {journal} {New Journal of Physics}\ }\textbf
  {\bibinfo {volume} {12}},\ \bibinfo {pages} {043048} (\bibinfo {year}
  {2010})}\BibitemShut {NoStop}%
\bibitem [{\citenamefont {Medhi}\ and\ \citenamefont
  {Shenoy}(2012)}]{Shenoy2012}%
  \BibitemOpen
  \bibfield  {author} {\bibinfo {author} {\bibfnamefont {Amal}\ \bibnamefont
  {Medhi}}\ and\ \bibinfo {author} {\bibfnamefont {Vijay~B}\ \bibnamefont
  {Shenoy}},\ }\bibfield  {title} {\enquote {\bibinfo {title} {Continuum theory
  of edge states of topological insulators: variational principle and boundary
  conditions},}\ }\href {http://stacks.iop.org/0953-8984/24/i=35/a=355001}
  {\bibfield  {journal} {\bibinfo  {journal} {Journal of Physics: Condensed
  Matter}\ }\textbf {\bibinfo {volume} {24}},\ \bibinfo {pages} {355001}
  (\bibinfo {year} {2012})}\BibitemShut {NoStop}%
\bibitem [{\citenamefont {Shen}\ \emph {et~al.}(2011)\citenamefont {Shen},
  \citenamefont {Shan},\ and\ \citenamefont {Lu}}]{SHEN2011}%
  \BibitemOpen
  \bibfield  {author} {\bibinfo {author} {\bibfnamefont {Shun-Qing}\
  \bibnamefont {Shen}}, \bibinfo {author} {\bibfnamefont {Wen-Yu}\ \bibnamefont
  {Shan}}, \ and\ \bibinfo {author} {\bibfnamefont {Hai-Zhou}\ \bibnamefont
  {Lu}},\ }\bibfield  {title} {\enquote {\bibinfo {title} {Topological
  insulator and the dirac equation},}\ }\href {\doibase
  10.1142/S2010324711000057} {\bibfield  {journal} {\bibinfo  {journal} {SPIN}\
  }\textbf {\bibinfo {volume} {01}},\ \bibinfo {pages} {33--44} (\bibinfo
  {year} {2011})}\BibitemShut {NoStop}%
\bibitem [{\citenamefont {Ryu}\ \emph {et~al.}(2010)\citenamefont {Ryu},
  \citenamefont {Schnyder}, \citenamefont {Furusaki},\ and\ \citenamefont
  {Ludwig}}]{Ryu2010}%
  \BibitemOpen
  \bibfield  {author} {\bibinfo {author} {\bibfnamefont {Shinsei}\ \bibnamefont
  {Ryu}}, \bibinfo {author} {\bibfnamefont {Andreas~P}\ \bibnamefont
  {Schnyder}}, \bibinfo {author} {\bibfnamefont {Akira}\ \bibnamefont
  {Furusaki}}, \ and\ \bibinfo {author} {\bibfnamefont {Andreas W~W}\
  \bibnamefont {Ludwig}},\ }\bibfield  {title} {\enquote {\bibinfo {title}
  {Topological insulators and superconductors: tenfold way and dimensional
  hierarchy},}\ }\href {http://stacks.iop.org/1367-2630/12/i=6/a=065010}
  {\bibfield  {journal} {\bibinfo  {journal} {New Journal of Physics}\ }\textbf
  {\bibinfo {volume} {12}},\ \bibinfo {pages} {065010} (\bibinfo {year}
  {2010})}\BibitemShut {NoStop}%
\bibitem [{\citenamefont {Munkres}(2000)}]{munkres2000topology}%
  \BibitemOpen
  \bibfield  {author} {\bibinfo {author} {\bibfnamefont {J.R.}\ \bibnamefont
  {Munkres}},\ }\href {https://books.google.com/books?id=XjoZAQAAIAAJ} {\emph
  {\bibinfo {title} {Topology}}},\ Featured Titles for Topology Series\
  (\bibinfo  {publisher} {Prentice Hall, Incorporated},\ \bibinfo {year}
  {2000})\BibitemShut {NoStop}%
\bibitem [{\citenamefont {Bernevig}\ and\ \citenamefont
  {Hughes}(2013)}]{bernevig2013topological}%
  \BibitemOpen
  \bibfield  {author} {\bibinfo {author} {\bibfnamefont {B~Andrei}\
  \bibnamefont {Bernevig}}\ and\ \bibinfo {author} {\bibfnamefont {Taylor~L}\
  \bibnamefont {Hughes}},\ }\href@noop {} {\emph {\bibinfo {title} {Topological
  insulators and topological superconductors}}}\ (\bibinfo  {publisher}
  {Princeton university press},\ \bibinfo {year} {2013})\BibitemShut {NoStop}%
\bibitem [{\citenamefont {Agranovich}\ and\ \citenamefont
  {Ginzburg}(2013)}]{agranovich2013crystal}%
  \BibitemOpen
  \bibfield  {author} {\bibinfo {author} {\bibfnamefont {Vladimir~M}\
  \bibnamefont {Agranovich}}\ and\ \bibinfo {author} {\bibfnamefont {Vitaly}\
  \bibnamefont {Ginzburg}},\ }\href@noop {} {\emph {\bibinfo {title} {Crystal
  optics with spatial dispersion, and excitons}}},\ Vol.~\bibinfo {volume}
  {42}\ (\bibinfo  {publisher} {Springer Science \& Business Media},\ \bibinfo
  {year} {2013})\BibitemShut {NoStop}%
\bibitem [{\citenamefont {Rukhadze}\ and\ \citenamefont
  {Silin}(1961)}]{Silin2961}%
  \BibitemOpen
  \bibfield  {author} {\bibinfo {author} {\bibfnamefont {Anri~A}\ \bibnamefont
  {Rukhadze}}\ and\ \bibinfo {author} {\bibfnamefont {Viktor~P}\ \bibnamefont
  {Silin}},\ }\bibfield  {title} {\enquote {\bibinfo {title} {Electrodynamics
  of media with spatial dispersion},}\ }\href
  {http://stacks.iop.org/0038-5670/4/i=3/a=R05} {\bibfield  {journal} {\bibinfo
   {journal} {Soviet Physics Uspekhi}\ }\textbf {\bibinfo {volume} {4}},\
  \bibinfo {pages} {459} (\bibinfo {year} {1961})}\BibitemShut {NoStop}%
\bibitem [{\citenamefont {Hestenes}(2002)}]{Hestenes2002}%
  \BibitemOpen
  \bibfield  {author} {\bibinfo {author} {\bibfnamefont {David}\ \bibnamefont
  {Hestenes}},\ }\enquote {\bibinfo {title} {Point groups and space groups in
  geometric algebra},}\ in\ \href {\doibase 10.1007/978-1-4612-0089-5_1} {\emph
  {\bibinfo {booktitle} {Applications of Geometric Algebra in Computer Science
  and Engineering}}},\ \bibinfo {editor} {edited by\ \bibinfo {editor}
  {\bibfnamefont {Leo}\ \bibnamefont {Dorst}}, \bibinfo {editor} {\bibfnamefont
  {Chris}\ \bibnamefont {Doran}}, \ and\ \bibinfo {editor} {\bibfnamefont
  {Joan}\ \bibnamefont {Lasenby}}}\ (\bibinfo  {publisher} {Birkh{\"a}user
  Boston},\ \bibinfo {address} {Boston, MA},\ \bibinfo {year} {2002})\ pp.\
  \bibinfo {pages} {3--34}\BibitemShut {NoStop}%
\bibitem [{\citenamefont {Nye}(1985)}]{nye1985physical}%
  \BibitemOpen
  \bibfield  {author} {\bibinfo {author} {\bibfnamefont {John~Frederick}\
  \bibnamefont {Nye}},\ }\href@noop {} {\emph {\bibinfo {title} {Physical
  properties of crystals: their representation by tensors and matrices}}}\
  (\bibinfo  {publisher} {Oxford university press},\ \bibinfo {year}
  {1985})\BibitemShut {NoStop}%
\bibitem [{\citenamefont {Hall}(2015)}]{hall_lie_2015}%
  \BibitemOpen
  \bibfield  {author} {\bibinfo {author} {\bibfnamefont {Brian~C.}\
  \bibnamefont {Hall}},\ }\href {\doibase 10.1007/978-3-319-13467-3} {\emph
  {\bibinfo {title} {Lie {Groups}, {Lie} {Algebras}, and {Representations}}}},\
  \bibinfo {series} {Graduate {Texts} in {Mathematics}}, Vol.\ \bibinfo
  {volume} {222}\ (\bibinfo  {publisher} {Springer International Publishing},\
  \bibinfo {address} {Cham},\ \bibinfo {year} {2015})\BibitemShut {NoStop}%
\bibitem [{\citenamefont {Fang}\ \emph {et~al.}(2012)\citenamefont {Fang},
  \citenamefont {Gilbert},\ and\ \citenamefont {Bernevig}}]{Fang2012}%
  \BibitemOpen
  \bibfield  {author} {\bibinfo {author} {\bibfnamefont {Chen}\ \bibnamefont
  {Fang}}, \bibinfo {author} {\bibfnamefont {Matthew~J.}\ \bibnamefont
  {Gilbert}}, \ and\ \bibinfo {author} {\bibfnamefont {B.~Andrei}\ \bibnamefont
  {Bernevig}},\ }\bibfield  {title} {\enquote {\bibinfo {title} {Bulk
  topological invariants in noninteracting point group symmetric insulators},}\
  }\href {\doibase 10.1103/PhysRevB.86.115112} {\bibfield  {journal} {\bibinfo
  {journal} {Phys. Rev. B}\ }\textbf {\bibinfo {volume} {86}},\ \bibinfo
  {pages} {115112} (\bibinfo {year} {2012})}\BibitemShut {NoStop}%
\bibitem [{\citenamefont {Heckenberg}\ \emph {et~al.}(1992)\citenamefont
  {Heckenberg}, \citenamefont {McDuff}, \citenamefont {Smith}, \citenamefont
  {Rubinsztein-Dunlop},\ and\ \citenamefont {Wegener}}]{Heckenberg1992}%
  \BibitemOpen
  \bibfield  {author} {\bibinfo {author} {\bibfnamefont {N.~R.}\ \bibnamefont
  {Heckenberg}}, \bibinfo {author} {\bibfnamefont {R.}~\bibnamefont {McDuff}},
  \bibinfo {author} {\bibfnamefont {C.~P.}\ \bibnamefont {Smith}}, \bibinfo
  {author} {\bibfnamefont {H.}~\bibnamefont {Rubinsztein-Dunlop}}, \ and\
  \bibinfo {author} {\bibfnamefont {M.~J.}\ \bibnamefont {Wegener}},\
  }\bibfield  {title} {\enquote {\bibinfo {title} {Laser beams with phase
  singularities},}\ }\href {\doibase 10.1007/BF01588597} {\bibfield  {journal}
  {\bibinfo  {journal} {Optical and Quantum Electronics}\ }\textbf {\bibinfo
  {volume} {24}},\ \bibinfo {pages} {S951--S962} (\bibinfo {year}
  {1992})}\BibitemShut {NoStop}%
\bibitem [{\citenamefont {Zeng}\ \emph {et~al.}(2016)\citenamefont {Zeng},
  \citenamefont {Zhu},\ and\ \citenamefont {Sheng}}]{Sheng2016}%
  \BibitemOpen
  \bibfield  {author} {\bibinfo {author} {\bibfnamefont {Tian-Sheng}\
  \bibnamefont {Zeng}}, \bibinfo {author} {\bibfnamefont {W.}~\bibnamefont
  {Zhu}}, \ and\ \bibinfo {author} {\bibfnamefont {D.~N.}\ \bibnamefont
  {Sheng}},\ }\bibfield  {title} {\enquote {\bibinfo {title} {Bosonic integer
  quantum hall states in topological bands with chern number two},}\ }\href
  {\doibase 10.1103/PhysRevB.93.195121} {\bibfield  {journal} {\bibinfo
  {journal} {Phys. Rev. B}\ }\textbf {\bibinfo {volume} {93}},\ \bibinfo
  {pages} {195121} (\bibinfo {year} {2016})}\BibitemShut {NoStop}%
\bibitem [{\citenamefont {Read}(1998)}]{Read1998}%
  \BibitemOpen
  \bibfield  {author} {\bibinfo {author} {\bibfnamefont {N.}~\bibnamefont
  {Read}},\ }\bibfield  {title} {\enquote {\bibinfo {title}
  {Lowest-landau-level theory of the quantum hall effect: The fermi-liquid-like
  state of bosons at filling factor one},}\ }\href {\doibase
  10.1103/PhysRevB.58.16262} {\bibfield  {journal} {\bibinfo  {journal} {Phys.
  Rev. B}\ }\textbf {\bibinfo {volume} {58}},\ \bibinfo {pages} {16262--16290}
  (\bibinfo {year} {1998})}\BibitemShut {NoStop}%
\bibitem [{\citenamefont {Zheng}\ and\ \citenamefont {Ando}(2002)}]{Zheng2002}%
  \BibitemOpen
  \bibfield  {author} {\bibinfo {author} {\bibfnamefont {Yisong}\ \bibnamefont
  {Zheng}}\ and\ \bibinfo {author} {\bibfnamefont {Tsuneya}\ \bibnamefont
  {Ando}},\ }\bibfield  {title} {\enquote {\bibinfo {title} {Hall conductivity
  of a two-dimensional graphite system},}\ }\href {\doibase
  10.1103/PhysRevB.65.245420} {\bibfield  {journal} {\bibinfo  {journal} {Phys.
  Rev. B}\ }\textbf {\bibinfo {volume} {65}},\ \bibinfo {pages} {245420}
  (\bibinfo {year} {2002})}\BibitemShut {NoStop}%
\bibitem [{\citenamefont {Jotzu}\ \emph {et~al.}(2014)\citenamefont {Jotzu},
  \citenamefont {Messer}, \citenamefont {Desbuquois}, \citenamefont {Lebrat},
  \citenamefont {Uehlinger}, \citenamefont {Greif},\ and\ \citenamefont
  {Esslinger}}]{Jotzu2014}%
  \BibitemOpen
  \bibfield  {author} {\bibinfo {author} {\bibfnamefont {Gregor}\ \bibnamefont
  {Jotzu}}, \bibinfo {author} {\bibfnamefont {Michael}\ \bibnamefont {Messer}},
  \bibinfo {author} {\bibfnamefont {R{\'e}mi}\ \bibnamefont {Desbuquois}},
  \bibinfo {author} {\bibfnamefont {Martin}\ \bibnamefont {Lebrat}}, \bibinfo
  {author} {\bibfnamefont {Thomas}\ \bibnamefont {Uehlinger}}, \bibinfo
  {author} {\bibfnamefont {Daniel}\ \bibnamefont {Greif}}, \ and\ \bibinfo
  {author} {\bibfnamefont {Tilman}\ \bibnamefont {Esslinger}},\ }\bibfield
  {title} {\enquote {\bibinfo {title} {Experimental realization of the
  topological haldane model with ultracold fermions},}\ }\href
  {http://dx.doi.org/10.1038/nature13915} {\bibfield  {journal} {\bibinfo
  {journal} {Nature}\ }\textbf {\bibinfo {volume} {515}},\ \bibinfo {pages}
  {237 EP --} (\bibinfo {year} {2014})}\BibitemShut {NoStop}%
\bibitem [{\citenamefont {Takayama}\ \emph {et~al.}(2008)\citenamefont
  {Takayama}, \citenamefont {Crasovan}, \citenamefont {Johansen}, \citenamefont
  {Mihalache}, \citenamefont {Artigas},\ and\ \citenamefont
  {Torner}}]{Takayama2008}%
  \BibitemOpen
  \bibfield  {author} {\bibinfo {author} {\bibfnamefont {Osamu}\ \bibnamefont
  {Takayama}}, \bibinfo {author} {\bibfnamefont {Lucian-Cornel}\ \bibnamefont
  {Crasovan}}, \bibinfo {author} {\bibfnamefont {Steffen~Kjær}\ \bibnamefont
  {Johansen}}, \bibinfo {author} {\bibfnamefont {Dumitru}\ \bibnamefont
  {Mihalache}}, \bibinfo {author} {\bibfnamefont {David}\ \bibnamefont
  {Artigas}}, \ and\ \bibinfo {author} {\bibfnamefont {Lluis}\ \bibnamefont
  {Torner}},\ }\bibfield  {title} {\enquote {\bibinfo {title} {Dyakonov surface
  waves: A review},}\ }\href {\doibase 10.1080/02726340801921403} {\bibfield
  {journal} {\bibinfo  {journal} {Electromagnetics}\ }\textbf {\bibinfo
  {volume} {28}},\ \bibinfo {pages} {126--145} (\bibinfo {year}
  {2008})}\BibitemShut {NoStop}%
\bibitem [{\citenamefont {Halevi}\ and\ \citenamefont
  {Fuchs}(1984)}]{Halevi1984}%
  \BibitemOpen
  \bibfield  {author} {\bibinfo {author} {\bibfnamefont {P}~\bibnamefont
  {Halevi}}\ and\ \bibinfo {author} {\bibfnamefont {R}~\bibnamefont {Fuchs}},\
  }\bibfield  {title} {\enquote {\bibinfo {title} {Generalised additional
  boundary condition for non-local dielectrics. i. reflectivity},}\ }\href
  {http://stacks.iop.org/0022-3719/17/i=21/a=017} {\bibfield  {journal}
  {\bibinfo  {journal} {Journal of Physics C: Solid State Physics}\ }\textbf
  {\bibinfo {volume} {17}},\ \bibinfo {pages} {3869} (\bibinfo {year}
  {1984})}\BibitemShut {NoStop}%
\bibitem [{\citenamefont {Gelfand}\ \emph {et~al.}(2000)\citenamefont
  {Gelfand}, \citenamefont {Fomin},\ and\ \citenamefont
  {Silverman}}]{gelfand2000calculus}%
  \BibitemOpen
  \bibfield  {author} {\bibinfo {author} {\bibfnamefont {I.M.}\ \bibnamefont
  {Gelfand}}, \bibinfo {author} {\bibfnamefont {S.V.}\ \bibnamefont {Fomin}}, \
  and\ \bibinfo {author} {\bibfnamefont {R.A.}\ \bibnamefont {Silverman}},\
  }\href {https://books.google.com/books?id=YkFLGQeGRw4C} {\emph {\bibinfo
  {title} {Calculus of Variations}}},\ Dover Books on Mathematics\ (\bibinfo
  {publisher} {Dover Publications},\ \bibinfo {year} {2000})\BibitemShut
  {NoStop}%
\bibitem [{\citenamefont {Hatsugai}(1993)}]{Hatsugai1993}%
  \BibitemOpen
  \bibfield  {author} {\bibinfo {author} {\bibfnamefont {Yasuhiro}\
  \bibnamefont {Hatsugai}},\ }\bibfield  {title} {\enquote {\bibinfo {title}
  {Chern number and edge states in the integer quantum hall effect},}\ }\href
  {\doibase 10.1103/PhysRevLett.71.3697} {\bibfield  {journal} {\bibinfo
  {journal} {Phys. Rev. Lett.}\ }\textbf {\bibinfo {volume} {71}},\ \bibinfo
  {pages} {3697--3700} (\bibinfo {year} {1993})}\BibitemShut {NoStop}%
\bibitem [{\citenamefont {Avila}\ \emph {et~al.}(2013)\citenamefont {Avila},
  \citenamefont {Schulz-Baldes},\ and\ \citenamefont
  {Villegas-Blas}}]{avila_topological_2013}%
  \BibitemOpen
  \bibfield  {author} {\bibinfo {author} {\bibfnamefont {Julio~Cesar}\
  \bibnamefont {Avila}}, \bibinfo {author} {\bibfnamefont {Hermann}\
  \bibnamefont {Schulz-Baldes}}, \ and\ \bibinfo {author} {\bibfnamefont
  {Carlos}\ \bibnamefont {Villegas-Blas}},\ }\bibfield  {title} {\enquote
  {\bibinfo {title} {Topological {Invariants} of {Edge} {States} for {Periodic}
  {Two}-{Dimensional} {Models}},}\ }\href {\doibase 10.1007/s11040-012-9123-9}
  {\bibfield  {journal} {\bibinfo  {journal} {Mathematical Physics, Analysis
  and Geometry}\ }\textbf {\bibinfo {volume} {16}},\ \bibinfo {pages}
  {137--170} (\bibinfo {year} {2013})}\BibitemShut {NoStop}%
\bibitem [{\citenamefont {Nagaosa}\ \emph {et~al.}(2010)\citenamefont
  {Nagaosa}, \citenamefont {Sinova}, \citenamefont {Onoda}, \citenamefont
  {MacDonald},\ and\ \citenamefont {Ong}}]{Nagaosa2010}%
  \BibitemOpen
  \bibfield  {author} {\bibinfo {author} {\bibfnamefont {Naoto}\ \bibnamefont
  {Nagaosa}}, \bibinfo {author} {\bibfnamefont {Jairo}\ \bibnamefont {Sinova}},
  \bibinfo {author} {\bibfnamefont {Shigeki}\ \bibnamefont {Onoda}}, \bibinfo
  {author} {\bibfnamefont {A.~H.}\ \bibnamefont {MacDonald}}, \ and\ \bibinfo
  {author} {\bibfnamefont {N.~P.}\ \bibnamefont {Ong}},\ }\bibfield  {title}
  {\enquote {\bibinfo {title} {Anomalous hall effect},}\ }\href {\doibase
  10.1103/RevModPhys.82.1539} {\bibfield  {journal} {\bibinfo  {journal} {Rev.
  Mod. Phys.}\ }\textbf {\bibinfo {volume} {82}},\ \bibinfo {pages}
  {1539--1592} (\bibinfo {year} {2010})}\BibitemShut {NoStop}%
\bibitem [{\citenamefont {Haldane}(1988)}]{Haldane1988}%
  \BibitemOpen
  \bibfield  {author} {\bibinfo {author} {\bibfnamefont {F.~D.~M.}\
  \bibnamefont {Haldane}},\ }\bibfield  {title} {\enquote {\bibinfo {title}
  {Model for a quantum hall effect without landau levels: Condensed-matter
  realization of the "parity anomaly"},}\ }\href {\doibase
  10.1103/PhysRevLett.61.2015} {\bibfield  {journal} {\bibinfo  {journal}
  {Phys. Rev. Lett.}\ }\textbf {\bibinfo {volume} {61}},\ \bibinfo {pages}
  {2015--2018} (\bibinfo {year} {1988})}\BibitemShut {NoStop}%
\bibitem [{\citenamefont {Fradkin}\ \emph {et~al.}(1986)\citenamefont
  {Fradkin}, \citenamefont {Dagotto},\ and\ \citenamefont
  {Boyanovsky}}]{Fradkin1986}%
  \BibitemOpen
  \bibfield  {author} {\bibinfo {author} {\bibfnamefont {Eduardo}\ \bibnamefont
  {Fradkin}}, \bibinfo {author} {\bibfnamefont {Elbio}\ \bibnamefont
  {Dagotto}}, \ and\ \bibinfo {author} {\bibfnamefont {Daniel}\ \bibnamefont
  {Boyanovsky}},\ }\bibfield  {title} {\enquote {\bibinfo {title} {Physical
  realization of the parity anomaly in condensed matter physics},}\ }\href
  {\doibase 10.1103/PhysRevLett.57.2967} {\bibfield  {journal} {\bibinfo
  {journal} {Phys. Rev. Lett.}\ }\textbf {\bibinfo {volume} {57}},\ \bibinfo
  {pages} {2967--2970} (\bibinfo {year} {1986})}\BibitemShut {NoStop}%
\bibitem [{\citenamefont {Shekhar}\ \emph {et~al.}(2017)\citenamefont
  {Shekhar}, \citenamefont {Malac}, \citenamefont {Gaind}, \citenamefont
  {Dalili}, \citenamefont {Meldrum},\ and\ \citenamefont
  {Jacob}}]{Shekhar2017}%
  \BibitemOpen
  \bibfield  {author} {\bibinfo {author} {\bibfnamefont {Prashant}\
  \bibnamefont {Shekhar}}, \bibinfo {author} {\bibfnamefont {Marek}\
  \bibnamefont {Malac}}, \bibinfo {author} {\bibfnamefont {Vaibhav}\
  \bibnamefont {Gaind}}, \bibinfo {author} {\bibfnamefont {Neda}\ \bibnamefont
  {Dalili}}, \bibinfo {author} {\bibfnamefont {Al}~\bibnamefont {Meldrum}}, \
  and\ \bibinfo {author} {\bibfnamefont {Zubin}\ \bibnamefont {Jacob}},\
  }\bibfield  {title} {\enquote {\bibinfo {title} {Momentum-resolved electron
  energy loss spectroscopy for mapping the photonic density of states},}\
  }\href {\doibase 10.1021/acsphotonics.7b00103} {\bibfield  {journal}
  {\bibinfo  {journal} {ACS Photonics}\ }\textbf {\bibinfo {volume} {4}},\
  \bibinfo {pages} {1009--1014} (\bibinfo {year} {2017})}\BibitemShut {NoStop}%
\bibitem [{\citenamefont {Scholl}\ \emph {et~al.}(2012)\citenamefont {Scholl},
  \citenamefont {Koh},\ and\ \citenamefont {Dionne}}]{scholl_quantum_2012}%
  \BibitemOpen
  \bibfield  {author} {\bibinfo {author} {\bibfnamefont {Jonathan~A.}\
  \bibnamefont {Scholl}}, \bibinfo {author} {\bibfnamefont {Ai~Leen}\
  \bibnamefont {Koh}}, \ and\ \bibinfo {author} {\bibfnamefont {Jennifer~A.}\
  \bibnamefont {Dionne}},\ }\bibfield  {title} {\enquote {\bibinfo {title}
  {Quantum plasmon resonances of individual metallic nanoparticles},}\ }\href
  {http://dx.doi.org/10.1038/nature10904} {\bibfield  {journal} {\bibinfo
  {journal} {Nature}\ }\textbf {\bibinfo {volume} {483}},\ \bibinfo {pages}
  {421} (\bibinfo {year} {2012})}\BibitemShut {NoStop}%
\bibitem [{\citenamefont {Saffman}\ \emph {et~al.}(2010)\citenamefont
  {Saffman}, \citenamefont {Walker},\ and\ \citenamefont
  {M\o{}lmer}}]{Saffman2010}%
  \BibitemOpen
  \bibfield  {author} {\bibinfo {author} {\bibfnamefont {M.}~\bibnamefont
  {Saffman}}, \bibinfo {author} {\bibfnamefont {T.~G.}\ \bibnamefont {Walker}},
  \ and\ \bibinfo {author} {\bibfnamefont {K.}~\bibnamefont {M\o{}lmer}},\
  }\bibfield  {title} {\enquote {\bibinfo {title} {Quantum information with
  rydberg atoms},}\ }\href {\doibase 10.1103/RevModPhys.82.2313} {\bibfield
  {journal} {\bibinfo  {journal} {Rev. Mod. Phys.}\ }\textbf {\bibinfo {volume}
  {82}},\ \bibinfo {pages} {2313--2363} (\bibinfo {year} {2010})}\BibitemShut
  {NoStop}%
\bibitem [{\citenamefont {Goban}\ \emph {et~al.}(2015)\citenamefont {Goban},
  \citenamefont {Hung}, \citenamefont {Hood}, \citenamefont {Yu}, \citenamefont
  {Muniz}, \citenamefont {Painter},\ and\ \citenamefont {Kimble}}]{Kimble2015}%
  \BibitemOpen
  \bibfield  {author} {\bibinfo {author} {\bibfnamefont {A.}~\bibnamefont
  {Goban}}, \bibinfo {author} {\bibfnamefont {C.-L.}\ \bibnamefont {Hung}},
  \bibinfo {author} {\bibfnamefont {J.~D.}\ \bibnamefont {Hood}}, \bibinfo
  {author} {\bibfnamefont {S.-P.}\ \bibnamefont {Yu}}, \bibinfo {author}
  {\bibfnamefont {J.~A.}\ \bibnamefont {Muniz}}, \bibinfo {author}
  {\bibfnamefont {O.}~\bibnamefont {Painter}}, \ and\ \bibinfo {author}
  {\bibfnamefont {H.~J.}\ \bibnamefont {Kimble}},\ }\bibfield  {title}
  {\enquote {\bibinfo {title} {Superradiance for atoms trapped along a photonic
  crystal waveguide},}\ }\href {\doibase 10.1103/PhysRevLett.115.063601}
  {\bibfield  {journal} {\bibinfo  {journal} {Phys. Rev. Lett.}\ }\textbf
  {\bibinfo {volume} {115}},\ \bibinfo {pages} {063601} (\bibinfo {year}
  {2015})}\BibitemShut {NoStop}%
\end{thebibliography}%

\end{document}